\documentclass[sn-mathphys-ay,pdflatex]{sn-jnl}

\usepackage{graphicx}%
\usepackage{amsmath,amssymb,amsfonts}%
\usepackage{array,multirow}
\usepackage{amsthm}%
\usepackage{mathrsfs}%
\usepackage{placeins}%
\usepackage[title]{appendix}%
\usepackage{xcolor}%
\usepackage{textcomp}%
\usepackage{manyfoot}%
\usepackage{booktabs}%
\usepackage{algorithm}%
\usepackage{algorithmicx}%
\usepackage{algpseudocode}%
\usepackage{listings}%

\usepackage{enumitem}

\newcommand{\gls}[1]{\MakeUppercase{#1}}
\newcommand{\glspl}[1]{\MakeUppercase{#1}}
\newcommand{\acrshort}[1]{\MakeUppercase{#1}}

\newcommand*\arcsec{\ensuremath{^{\prime\prime}}}
\newcommand*\farcs{\ensuremath{\overset{\prime\prime}{.}}}

\newcommand{\T}{\mathsf{T}} 

\newcommand{\MtoC}{\mathbf{M2C}} 

\newcommand{\N}{\mathbf{N}} 
\newcommand{\D}{\mathbf{W}} 

\newcommand{\VDM}{\mathsf{{VDM}}} 

\newcommand{\Mfour}{\mathsf{{M4}}}

\newcommand{\svec}{{\mathbf{s}}}
\newcommand{\mvec}{{\mathbf{m}}}

\newcommand{\evec}{{\mathbf{e}}} 
\newcommand{\uvec}{{\mathbf{u}}}

\newcommand{\mum}{\ensuremath{\mu\mathrm{m}}}

\begin{document}

\title{High Strehl and High Contrast for the ELT Instrument METIS} 
\subtitle{Final Design, Implementation, and Predicted Performance of the Single-Conjugate Adaptive Optics System}

\author[1]{\fnm{Markus}\sur{Feldt}\email{feldt@mpia.de}}
\author[1]{\fnm{Thomas}\sur{Bertram}\email{bertram@mpia.de}}
\author[2]{\fnm{Carlos}\sur{Correia}\email{carlos.correia@fe.up.pt}}
\author[3]{\fnm{Olivier}\sur{Absil}\email{olivier.absil@uliege.be}}
\author[1]{\fnm{M. Concepción}\sur{Cárdenas Vázquez}\email{conchi@mpia.de}}
\author[1]{\fnm{Hugo}\sur{Coppejans}\email{coppejans@mpia.de}}
\author[1]{\fnm{Martin}\sur{Kulas}\email{kulas@mpia.de}}
\author[4]{\fnm{Andreas}\sur{Obereder}\email{andreas.obereder@mathconsult.co.at}}
\author[3]{\fnm{Gilles}\sur{Orban de Xivry}\email{gorban@uliege.be}}
\author[1]{\fnm{Silvia}\sur{Scheithauer}\email{scheithauer@mpia.de}}
\author[1]{\fnm{Horst}\sur{Steuer}\email{steuer@mpia.de}}

\affil*[1]{\orgname{Max Planck Institute for Astronomy}, \orgaddress{\street{Königstuhl 17}, \postcode{D-69117} \city{Heidelberg}, \country{Germany}}}

\affil*[2]{\orgdiv{Faculdade de Engenharia da Universidade do Porto}, \orgname{Universidade do Porto}, \orgaddress{\street{Rua Dr. Roberto Frias}, \postcode{Pt-4200-465} \city{Porto}, \country{Portugal}}}

\affil*[3]{\orgdiv{STAR Institute}, \orgname{Université de Liège}, \orgaddress{\street{Allée du six Août 19c}, \postcode{B-4000} \city{Liège}, \country{Belgium}}}

\affil*[4]{\orgdiv{Institute for Industrial Mathematics}, \orgname{Linz University}, \orgaddress{\street{Altenbergerstraße 69}, \postcode{A-4040} \city{Linz}, \country{Austria}}}


\abstract{The Mid-infrared ELT Imager and Spectrograph (\gls{metis}) is a first-generation instrument for the Extremely Large Telescope (\gls{elt}), Europe’s next-generation 39\,m ground-based telescope for optical and infrared wavelengths, which is currently under construction at the European Southern Observatory (\gls{eso}) site at Cerro Armazones in Chile.

METIS will offer diffraction-limited imaging, low- and medium-resolution slit spectroscopy, and coronagraphy for high-contrast imaging between 3 and 13 microns, as well as high-resolution integral field spectroscopy between 3 and 5 microns.
The main METIS science goals are the detection and characterisation of exoplanets, the investigation of proto-planetary disks, and the formation of planets.

The Single-Conjugate Adaptive Optics (\gls{scao}) system corrects atmospheric distortions and is thus essential for diffraction-limited observations with METIS. \gls{scao} will be used for all observing modes, with high-contrast imaging imposing the most demanding requirements on its performance.

The Final Design Review (\gls{fdr}) of METIS took place in the fall of 2022; the development of the instrument, including its \gls{scao} system, has since entered the Manufacturing, Assembly, Integration and Testing (\gls{mait}) phase. 
Numerous challenging aspects of an \gls{elt} Adaptive Optics (\gls{ao}) system are addressed in the mature designs for the \gls{scao} control system and the \gls{scao} hardware module: the complex interaction with the telescope entities that participate in the AO control, wavefront reconstruction with a fragmented and moving pupil, secondary control tasks to deal with differential image motion, non-common path aberrations and mis-registration. 
A \textit{K}-band pyramid wavefront sensor and a GPU-based Real-Time Computer (\gls{rtc}), tailored to the needs of \gls{metis} at the \gls{elt}, are core components. 


This current paper serves as a natural sequel to our previous work presented in \cite{Hippler18b}. It reflects all the updates that were implemented between the Preliminary Design Review (\gls{pdr}) and \gls{fdr}, and includes updated performance estimations in terms of several key performance indicators, including achieved contrast curves.  We outline all important design decisions that were taken, and present the major challenges we faced and the main analyses carried out to arrive at these decisions and eventually the final design.  We also elaborate on our testing and verification strategy, and, last not least, comprehensively present the full design, hardware and software in this paper to provide a single source of reference which will remain valid at least until commissioning.
}




\maketitle




\section{Overview}

The \gls{metis} science goals set stringent requirements on its scientific sub-systems, as well as its \gls{scao} sub-system. In particular the exoplanet science case demands exquisite high-contrast performance, which in turn requires a degree of control of the wavefront that goes beyond the usual Strehl thresholds, in particular when combined with the fragmented nature of the ELT pupil. 

\gls{scao} itself is a distributed system, where the \gls{scao} module, which comprises essentially the Wave Front Sensor (\gls{wfs}), is located inside the instrument itself where it  measures the disturbed incoming wavefront.  This \gls{wfs} location inside the main cryostat was necessary due to the nature of METIS being a Mid-Infrared (\gls{mir}) instrument that cannot tolerate a warm beam-splitter surface between the \gls{wfs} and the science paths.
The \gls{wfs} signal is processed by the \gls{rtc}, which combines software and hardware components and communicates with the Central Control System (\gls{ccs}) of the \gls{elt}.  The \gls{ccs} then steers the adaptive elements within the \gls{elt} optical train, among them the high-order deformable mirror M4, and the tip-tilt mirror M5.

A novel approach to wavefront control is deployed, in which the reconstruction of the wavefront from the sensor signals is logically separated from the projection onto the deformable mirror control modes and the temporal filtering. In this approach, a virtual deformable mirror is introduced, which is kept fixed in orientation with respect to the wavefront sensor and serves as an intermediate layer in which the wavefront is first reconstructed. In a second step, the alignment of the wavefront sensor (and thus the virtual mirror) with the actual M4 is taken into account and the wavefront is projected onto a dedicated set of mirror control modes, and finally the corresponding command signals. 

A main limitation for the performance of \gls{metis} and its \gls{scao} system are Non-Common Path Aberrations (\gls{ncpa}s) and water vapor seeing.  In order to at least partially overcome this limitation, we foresee focal plane wavefront sensing, where image data from the science focal planes are made available to the \gls{rtc} and dedicated algorithms are deployed to measure residual aberrations in the coronagraphic and/or science focal plane, which can then be corrected by means of offsets applied to the measured slopes measured by the \gls{wfs}. 

In this paper, we first describe the main characteristics of \gls{metis} itself, its science cases and the resulting requirements in Sec.~\ref{sec:METIS}. We outline a series of design decisions that were taken in the course of devising the \gls{scao} system in order to enable it meeting all challenges.  

We will then describe in detail the hardware design of the \gls{scao} module and our integration and testing strategy in Sec.~\ref{sec:scao_module_design}. 

In Sec.~\ref{sec:wfc-strategy} we describe our wavefront control strategy split in the main control loop, auxiliary loop and loop co-processing tasks.

Sec.~\ref{sec:rtc} describes the RTC hard- and software design, the current status, and the tests already carried out and still foreseen.

The predicted performance of \gls{metis}' \gls{scao} system as assessed in numerous simulations is presented in Sec.~\ref{sec:scao-sim}, along with detailed analyses regarding \gls{scao}'s behaviour in the presence of \gls{ncpa}s, the low-wind phenomenon, and a combination of real-world adversarial effects.

Finally, last but not least Sect.~\ref{sec:hci} presents detailed simulations of the expected high-contrast performance of the entire \gls{metis} instrument.

Several aspects treated in this article have been highlighted in conference proceedings before (\cite{bertram23}; \cite{kulas23}; \cite{correia22}; \cite{feldt23}). Although these remain excellent sources for in-depth information on their respective topics, they were generally based on pre-\gls{fdr} analyses and results.  The design and predicted performance of METIS SCAO is now unlikely to change throughout the \gls{mait} and the subsequent commissioning phases, so this paper as a comprehensive update to \cite{Hippler18b}, will see its next update once commissioning results become available.


\section{METIS, a science-led ELT 1st-light instrument \label{sec:METIS}}
\subsection{Science with METIS \label{sec:science_with_metis}}

\gls{metis} will be \textit{the} planet finder and characterisation instrument at the \gls{elt} for at least its first five years of operation. 
It will provide diffraction-limited spectroscopy and imaging, including coronagraphic capabilities,  in the thermal/mid-infrared wavelength domain (3 $\mu$m – 13.3 $\mu$m), with High-Contrast Imaging (\gls{hci}) imposing the most demanding requirements on its performance.

Among many science cases that will benefit from these new capabilities, two key cases have been identified that will drive the scientific requirements on the instrument: extrasolar planets and circumstellar disks.  Nevertheless, one should keep in mind that, unlike e.g. the SPHERE instrument \citep{beuzit19}, specialized for imaging and characterisation of exoplanets and in operation at \gls{eso}'s Very Large Telescope (\gls{vlt}), \gls{metis} will be a general-purpose tool for the astronomical community.  One can clearly expect that \gls{metis} will add key insights in all areas of mid-infrared astronomy where high spatial or high spectral resolution, or a combination of both, is crucial.  The \gls{metis} science case thus covers a wide field of topics, such as
\begin{itemize}
    \item  Protoplanetary Disks and the Formation of Extrasolar Planets
    \item   Exoplanet Detection and Characterisation
    \item   The Formation History of the Solar System
    \item   The Properties of Low-Mass Brown Dwarfs
    \item   Massive Stars and Cluster Formation
    \item   Evolved Stars and their Circumstellar Environment
    \item   The Galactic Center
    \item   Extragalactic Science such as Starbursts in the Local Universe, Luminous Star-forming Galaxies at high Redshift, Active Galactic Nuclei (AGNs), and Transient Events
\end{itemize}

\subsection{Instrumentation suite}

 {\renewcommand{\arraystretch}{1.2}
 \begin{table}[h!]
 \caption{METIS Primary Observing Modes. The HCI masks are the Ring-Apodized Vortex Coronagraph (RAVC), Classical Vortex Coronagraph (CVC), and the Apodized Phase Plate (APP). }
 \label{tab:obsmodes}
 \begin{tabular}{p{2cm}cp{1.5cm}ccp{3cm}}
 \toprule
 \multirow{2}{*}{Observing Mode} & \multicolumn{4}{c}{Instrument Config.} & \multirow{2}{*}{Science Case} \\ \cmidrule{2-5}

 & HCI Mask & Band & FOV & Spec. Res.  &  \\
 \midrule \midrule
\multirow{2}{*}{Direct Imaging} & \multirow{2}{*}{--} & $L$/$M$ &  10\farcs5$\times$10\farcs5 &  \multirow{2}{*}{6-70}&\multirow{2}{*}{\parbox{3cm}{\footnotesize circum-stellar / cir\-cum-nuclear structures, clusters}} \\ 
& & $N$ & 13\farcs5$\times$13\farcs5 & &  \\
\hline 

\multirow{2}{*}{\parbox{2cm}{High-Contrast Imaging}} & Any & $L$/$M$ &  10\farcs5$\times$10\farcs5 &  \multirow{2}{*}{6-70}&\multirow{2}{*}{\parbox{3cm}{exoplanets}} \\ 
& CVC & $N$ & 13\farcs5$\times$13\farcs5 & &  \\
\hline

\multirow{2}{*}{\parbox{2cm}{Long-Slit Spectroscopy}} & -- & $L$/$M$ & 10\farcs5 & 1400-1900 & Ices \\ 
                                                        & -- & $N$     & 13\farcs5 & 400 & physics, chemistry, mineralogy of circum-stellar and circum-nuclear environments \\ 
\hline

\multirow{2}{*}{\parbox{2cm}{IFU Spectroscopy}} & -- & $LM$, $\Delta\lambda=$40nm-60nm & 0\farcs58$\times$0\farcs90 & {100k} & kinematics and chemistry of circum-stellar environments \\ 
                                                & -- & $LM$, $\Delta\lambda=$300nm & 0\farcs062$\times$0\farcs90 & 100k & chemistry of interstellar medium \\

                                                 \hline
\multirow{2}{*}{\parbox{2cm}{IFU+HCI Spectroscopy}} & Any & $LM$, $\Delta\lambda=$40nm-60nm & 0\farcs58$\times$0\farcs90 & {100k} & kinematics and chemistry of circum-stellar environments \\ 
                                                & Any & $LM$, $\Delta\lambda=$300nm & 0\farcs062$\times$0\farcs90 & 100k & chemistry of interstellar medium \\

 \bottomrule
 \end{tabular}
 \end{table}
 }

METIS is a cryogenic instrument that includes an imaging science subsystem (IMG) with a sophisticated coronagraph for \gls{hci}, an LM band spectrograph (LMS), and the wavefront control sub-system \gls{scao}.

A suite of observing modes will offer
\begin{itemize}
\item direct imaging at L, M, and N band
\item \gls{hci} at L, M, and N band, deploying different coronagraphic techniques \citep{absil23}
\item longslit spectroscopy at L, M and N band with a resolution between 400 and 2000
\item high-resolution (R~$\sim~10^5$) integral-field spectroscopy at L/M band, including a mode with extended instantaneous wavelength coverage
\item simultaneous \gls{hci} and integral-field spectroscopy 
\end{itemize}

A summary of the modes can be found in Tab.~\ref{tab:obsmodes}

\begin{figure} [h!]
   \begin{center}
   \includegraphics[width=0.45\textwidth]{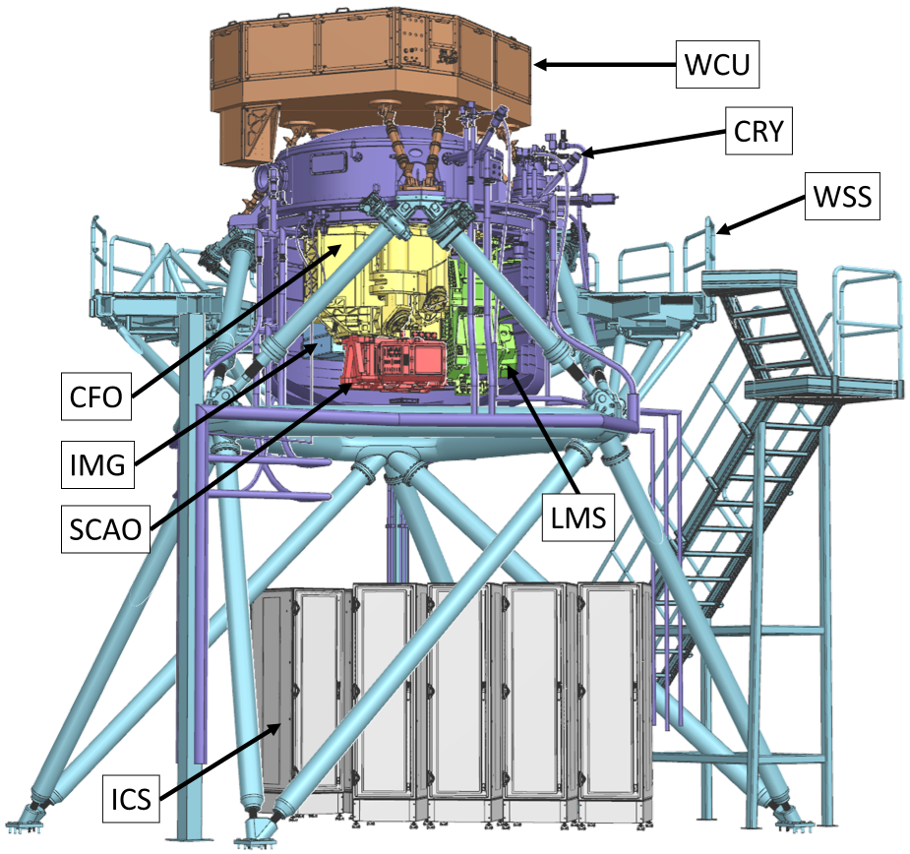}
   \end{center}
   \caption{METIS will be located on one of the Nasmyth platforms of the ELT.
   At a height of 6m above the platform, the light coming from the telescope enters the METIS cryostat, which hosts the entire optical path of the instrument down to the science detectors.
   It provides a very stable cryogenic environment, suppressing thermal background emission. The SCAO module also resides within the cryostat.
   Credit: METIS consortium}
\label{fig:metis_fdr_design}
\end{figure}

Thermal self-emission is by far the dominant noise contribution in the wavelength range where METIS operates. To limit its impact it is essential for the instrument to be realized within a stable cryogenic environment (cf. Figure~\ref{fig:metis_fdr_design}). The cryostat (CRY) has a diameter of 3\,m and is located on the \gls{elt} Nasmyth platform. It is held by a Warm Support Structure (WSS) and hosts the science subsystems imager (IMG) and LM-spectrograph (LMS), as well as the \gls{scao} module and the Common Fore Optics (CFO). On top of it, the Warm Calibration Unit (WCU) is shown. Underneath the cryostat the (warm) electronic cabinets (ICS) are located.

\gls{metis} is being built by an international consortium under the leadership of NOVA in the Netherlands. A dedicated overview on \gls{metis} is given in \cite{2021Msngr.182...22B, 2022SPIE12184E..21B}.
The development of its \gls{scao} system is led by the Max Planck Institute for Astronomy (\gls{mpia}) in Germany.

\subsection{Project Timeline}

The \gls{fdr} of \gls{metis} took place in the fall of 2022. The project is now in its \gls{mait} phase, where the individual subsystems are being manufactured, assembled, integrated, and tested by the consortium partners. Starting 2025, the sub-systems will subsequently come together at the main Assembly, Integration, and Testing (\gls{ait}) facility in Leiden, The Netherlands, where the system \gls{ait} will happen. \gls{scao} will be one of the last subsystems to be integrated at system level in 2027. It will have been extensively tested at subsystem level by then using a telescope simulator mimicking the \gls{elt} and thus allowing for \gls{ao} closed-loop testing, see Section~\ref{sec_testing}. The system tests shall finish in 2028 to have the Preliminary Acceptance Europe (PAE) review in summer 2028. After a successful PAE the instrument will be shipped to Chile and installed and commissioned at the \gls{elt}.

\subsection{SCAO Performance Requirements}

Each observing mode of \gls{metis} is designed to fully exploit the diffraction limit of the \gls{elt}'s 39m aperture.
\gls{metis} \gls{scao} is the real-time wavefront control system that enables diffraction-limited observations.
High-contrast imaging imposes challenging requirements on \gls{metis} \gls{scao}.
In order to reach the  (post processed) contrast level pursued (cf. Table \ref{tab:requirements}), the performance of the \gls{scao} system must not only be met in terms of the Strehl ratio.
Other parameters, such as the residual pointing jitter and the petal piston error, also need to be constrained.

\begin{table}[h!]
\caption{Selected top level requirements for METIS SCAO}
\label{tab:requirements}
\begin{tabular}{llll}
\toprule
              &  Requirement & Goal & $\lambda$\\
\midrule
Strehl Ratio & $>$ 93 \%     & $>$ 95 \% & 10 $\mu$m \\
              & $>$ 60 \%     & $>$ 80 \% & 3.7 $\mu$m \\
 & & &\\
 Contrast &  $<3 \cdot 10^{-5}$ & $< 10^{-6}$     & 3.7 $\mu$m\\
          &  at 5$\lambda$/D    & at 2$\lambda$/D &           \\
\bottomrule
\end{tabular}
\end{table}

\subsection{NCPA in METIS}\label{sec:HCI_NCPA}
As mentioned in Sec.~\ref{sec:science_with_metis} \textit{the} \gls{metis} science case is exoplanets.  Because they create long-lived speckles that can be confused with planetary signals, \gls{ncpa} are widely recognized to be one of the main limitations to the performance of both current and future \gls{hci} instruments. Even with dedicated sensing and compensation techniques deployed (c.f.\ Sec.~\ref{sec:design_decisions}), \gls{ncpa} are also expected to have an impact on the \gls{scao} performance, depending on how they are handled, and in turn again impact the \gls{hci} performance. This is specifically analysed in  Sec.~\ref{sec:HCI_NCPA_correction}.

Beyond static \gls{ncpa} - which can be measured during Assembly, Integration, and Verification (\gls{aiv}) - \gls{metis} will have two main contributions of varying \gls{ncpa}: chromatic beam wander, related to the changing position of the beam footprint on the METIS optical train during the observation, and water vapor seeing.
Indeed, differential atmospheric refraction leads to a changing relative position of the beam in $LMN$ bands compared to the fixed, \gls{scao}-stabilized $K$-band beam. A detailed discussion of this effect and how this is modelled for METIS can be found in \citet[][in prep.]{Bone+2024}.
The second contribution, water vapor seeing, is due to the non-uniform distribution of the water and the strong chromaticity of the water refractive index in the mid-infrared. The nonuniformity of water in the air creates additional spatial variations in the optical path, and the wind transports those variations over time following the frozen flow hypothesis. 
Since the wavefront is controlled at near-infrared wavelengths ($K$-band), this dispersion leads to strong additional wavefront errors at the scientific mid-infrared wavelengths (from $L-$ to $N-$ bands). 
We measured resulting optical path differences on the six Very Large Telescope Interferometer (\gls{vlti}) baselines to be equal to 50\,nm at $L$-band and 600nm at $N$-band. Assuming von Karman statistics, this results 
in wavefront rms of 30nm in $L$-band and $350$nm in $N$-band.
A thorough discussion of water vapour seeing effect and how it is modeled for METIS is presented in \cite{Absil+2022}.

\subsection{Design Decisions \label{sec:design_decisions}}

The design of the \gls{metis} \gls{scao} system is based on a number of choices made in earlier project phases and supported by dedicated studies.

\subsubsection{Wavefront sensing wavelength} The \gls{wfs} of \gls{metis} \gls{scao} operates in the K-band. 
Wavefront sensing in the H-band is also supported. These bands come with a number of advantages:
Using the atmospheric window next to the science wave bands minimizes the amount of chromatic aberration. 
Very efficient and mature NIR detector technology is available for wavefront sensing applications. 
And last but definitely not least: the longer wavefront sensing wavelength helps to overcome a difficulty, the so-called 'petaling effect', that is inherent to all "extremely" large telescopes. 
Their large support structures dissect the pupil images as seen by the wavefront sensors, causing problems to establish and maintain a phase relationship between the individual pupil fragments. 

\subsubsection{Type of wavefront sensor} A Pyramid Wave Front Sensor (\gls{pwfs}) \citep{1996JMOp...43..289R} with a spatial sampling of 90 $\times$ 90 subapertures has been developed for \gls{metis} \gls{scao}. 
The \gls{pwfs} provides the best combination of sensitivity, flexibility, but also maturity and is the type of \gls{wfs} that is used for the vast majority of Natural Guide Star (\gls{ngs})-fed \gls{ao} systems that are currently under construction.
Unlike the Shack-Hartmann \gls{wfs} the \gls{pwfs} allows us to deal with the petaling effect in that it preserves information on phase differences between neighbouring pupil fragments.

\subsubsection{Wavefront sensor in cryogenic environment} METIS' science wavelengths range from 3\,\mum{} to 13.3\,\mum.  It is thus of paramount importance to avoid that radiation emanating from  any kind of warm (i.e. at ambient temperature) surface reaches any of METIS' science detectors.  Early analyses of the thermal background showed that the beam splitter between the science path and the \gls{wfs} path needs to be within the cryogenic environment, and in order to avoid additional windows and complexity it was decided that the full \gls{wfs} module would also be located inside the cryostat.

\subsubsection{Numerical misregistration correction}
Pupil rotation, possibly inaccurate angular tracking of the platforms, but also flexure within the telescope precludes a static projection of the actuator positions onto the \gls{wfs}.
Additional actuators in the optical path could be introduced to counteract the motion of the image of the Deformable Mirror (\gls{dm}) relative to the \gls{wfs}. To keep the amount of active components in the cryogenic \gls{scao} module to a minimum, it was decided to instead consider misregistration between the ELT's deformable mirror \gls{m4} and the \gls{pwfs} through recurrent updates of the command matrix. 
This approach is very flexible, as it is not limited to the anticipated degrees of freedom that could have been foreseen in an opto-mechanical solution.

\subsubsection{Focal plane wave front sensing}
Due to the severe impact of fast changing \gls{ncpa}s (see Sec.~\ref{sec:HCI_NCPA}), additional wavefront sensing at the science wavelengths is preferential. This will be provided by a direct access of the \gls{metis} \gls{rtc} to focal plane data in near real-time, and the two auxiliary control loops using a technique called "Quadrant Analysis of Coronagraphic Images for Tip-tilt Sensing" (\gls{qacits}) to correct pointing errors through modulator offsets, and using an Asymmetric Lyot Wave Front Sensor (\gls{alwfs}) in combination with the Vortex coronagraphs to correct high-order \gls{ncpa}s  through WFS slope modifications.  The latter is a novel wavefront sensing approach using an asymmetric Lyot stop and machine learning, specifically developed to mitigate the impact of water vapour seeing by providing an instantaneous measurement based on focal plane images \citep[see][]{Orban+2024}. Similarly, the Apodising Phase Plate (\gls{app}) may also be combined with an asymmetric pupil to enable wavefront sensing (see e.g. \cite{Bos+2019}).

\begin{figure*} [h!]
   \begin{center}
   \includegraphics[width=0.66\textwidth]{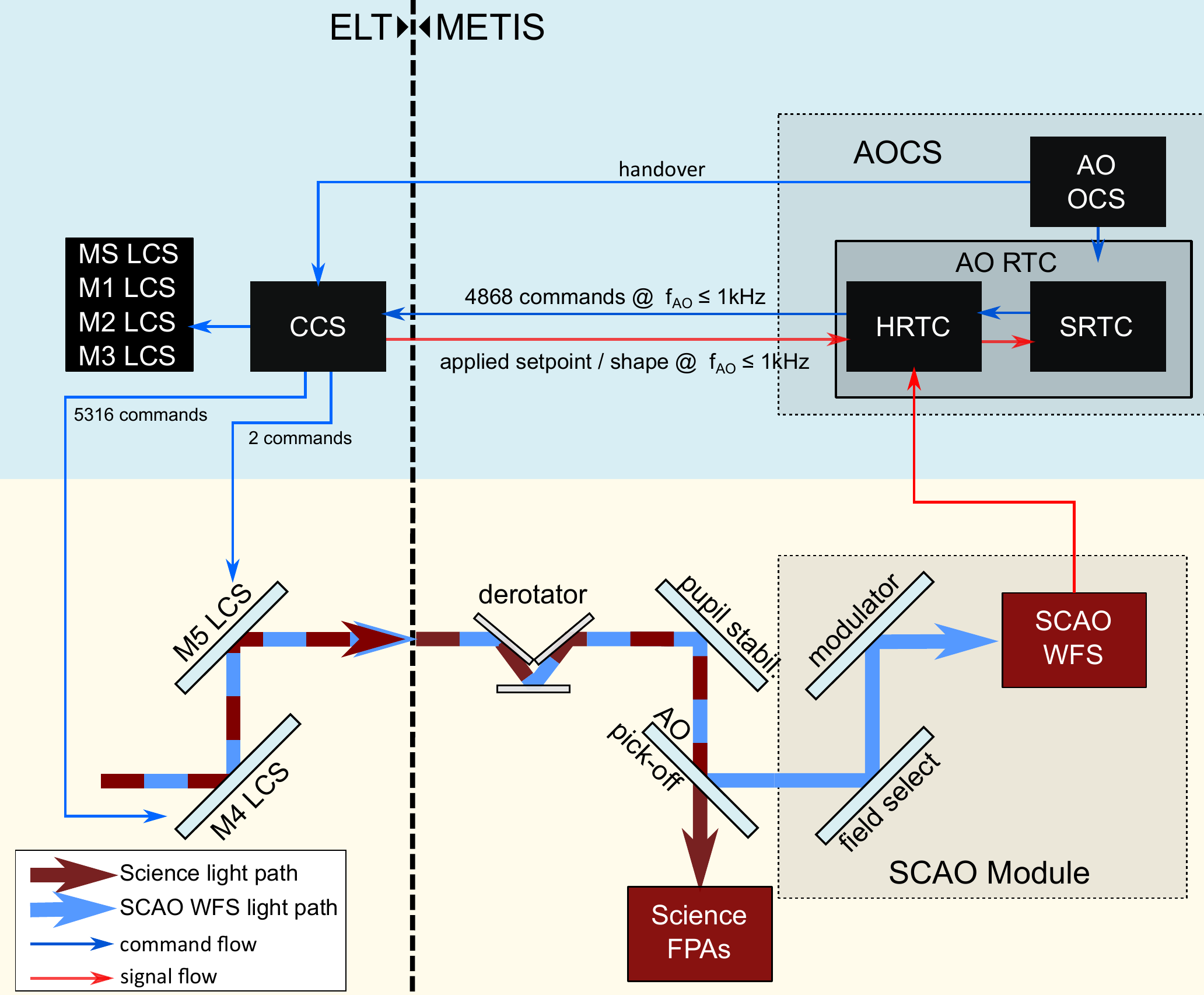}
   \end{center}
   \caption{Simplified block diagram of the METIS SCAO system: The AO Control System (\gls{aocs}) and the SCAO Module (shaded boxes) are the entities of the SCAO system that belong to the instrument domain.
   The key entities for the real-time correction of the incoming light of the ’ELT’ domain are located on the left side of the figure.
   In a closed wavefront control loop, the blue, NIR light is used to measure the instantaneous residual wavefront error by the wavefront sensor (SCAO WFS).
   The measurement signal is analysed by the Real-Time Computer (AO RTC), and a computed correction is sent to the Central Control System (CCS) to be applied with the adaptive mirrors M4 and M5 of the ELT. Taken from \cite{bertram23}}
\label{fig:block_diagram}
\end{figure*}

\subsection{Design of a Distributed SCAO System}

\begin{table}[h!]
\caption{SCAO System Parameters}
\begin{tabular}{rc}
\toprule
  Parameter & Value \\
\midrule

WFS Type                 & cryogenic pyramid WFS \\
Operation Wavelength & 1.45-2.45\,$\mu$m \\
Number of Subapertures & $90\times90$ \\
Max. Loop Freq.      & 1\,kHz \\ 
Modulation range     & 0-10 $\lambda/D$ (rad.) \\
Patrol Area Diam.    & 27\arcsec \\
Detector               & SAPHIRA \\
No. of Actuators        & 5352 + 2 \\
RTC technology         & GPU \\
\bottomrule
\end{tabular}
\label{tab:scao_parameters}
\end{table}

The wavefront control system for \gls{metis} is based on several distributed entities: the \emph{SCAO Module} and the \emph{AO Control System (AOCS)} in the instrument domain, as well as on the corrective optics (M4 and M5) and the \gls{ccs} in the telescope domain.
Figure \ref{fig:block_diagram} shows a simplified block diagram for \gls{metis} \gls{scao}.

The \emph{SCAO Module} is located inside the cryostat.
A cold dichroic pick-off mirror immediately in front of the \gls{scao} module is used to separate the Near-Infrared (\gls{nir}) part of the light, which is used for wavefront sensing.
The SCAO Module provides a \gls{pwfs} as well as optomechanical actuators for field selection and modulation of the natural guide star in the accessible field of view.

The \emph{AO Control System (AOCS)} hosts the main wavefront control loop as well as a number of secondary control loops.
A key entity of the \gls{aocs} is the \gls{rtc} \citep{kulas23}.
Its Hard Real-Time Core (HRTC) is used for the time critical aspects of the wavefront control loop: wavefront sensor signal processing, wavefront reconstruction, and the determination of correction commands that are applied with the adaptive M4 and M5 mirrors of the \gls{elt} via the \gls{ccs}.

The key parameters of the \gls{scao} system are summarized in Tab.~\ref{tab:scao_parameters}

\subsection{SCAO Development Plan}

\begin{figure*} [ht]
   \begin{center}
   \includegraphics[width=0.99\textwidth]{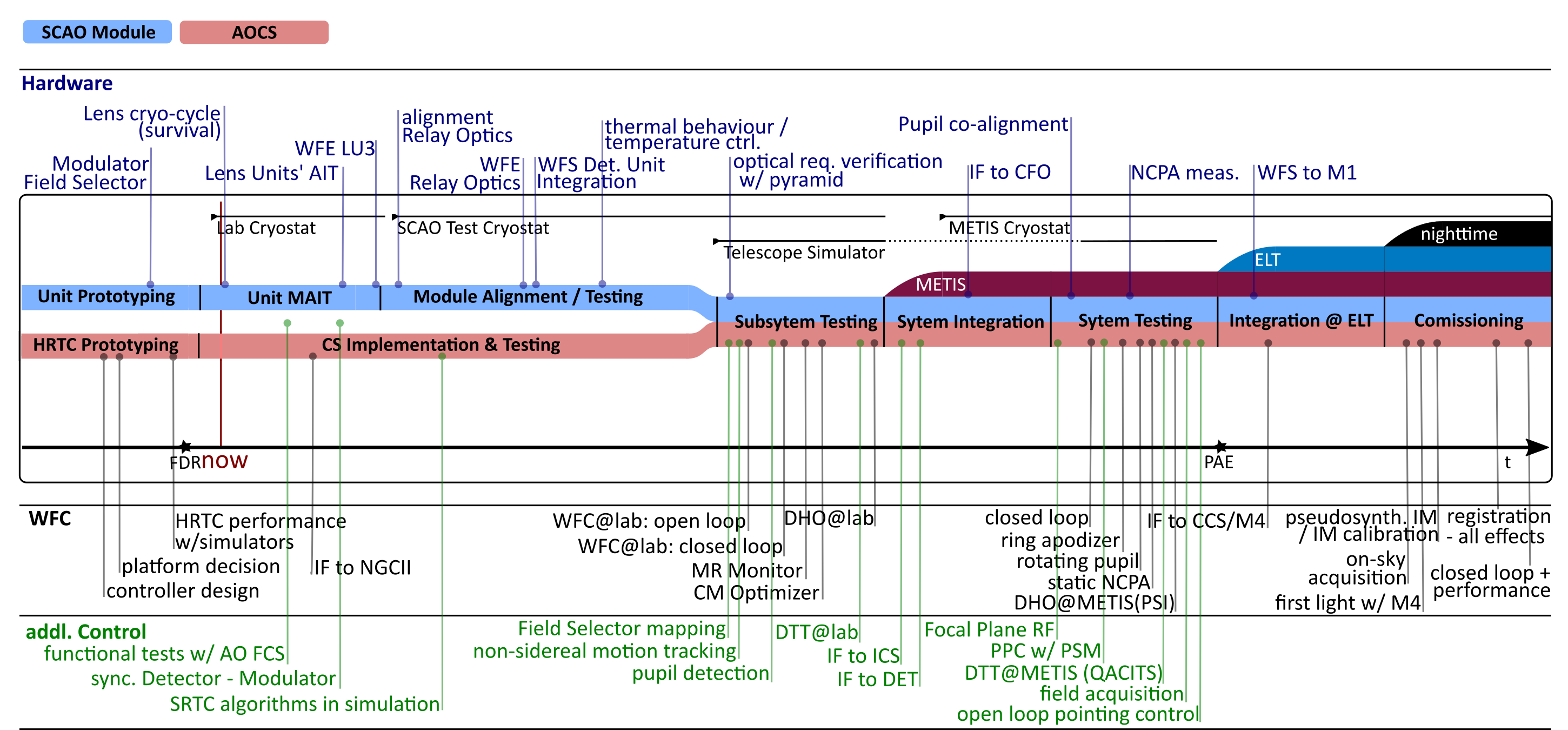}
   \end{center}
   \caption{MAIT road map: The development of the AO Control System (AOCS, red) and the SCAO Module (blue) is done in parallel. Each MAIT phase represents a higher level of integration. Dedicated test setups, such as the SCAO test cryostat or the telescope simulator will be used in the corresponding test phases. Key tests for the hardware, the main wavefront control (WFC) pipeline and for auxiliary control tasks are indicated.}
\label{fig:mait_flow} 
\end{figure*}

The completion of the final design allows \gls{metis} to proceed to the \gls{mait} phase.
\gls{scao} \gls{mait} activities span the entire remaining METIS development cycle until it is fully operational.
Many aspects, especially related to the \gls{aocs}, cannot be tested in the subsystem \gls{ait} phase of \gls{metis}.
Figure \ref{fig:mait_flow} provides a bird-eye view of the test activities and their dependencies on the test setups and the level of integration.
The first parallel and then joint development of the SCAO
Module and \gls{aocs} is shown.

The procurement of most optical components and mechanisms of the SCAO Module has a lead time of more than 6 months.
These components had been reviewed as part of a long-lead item \gls{fdr} before the system \gls{fdr} and are ready for integration.

As many of the core components of \gls{metis} \gls{scao} are part of the telescope domain and available for integration and testing only at a very late stage in the development, some only after the instrument has been installed at the \gls{elt}, it is important to establish environments that allow for thorough testing at much earlier stages.

The \gls{mait} plan foresees different phases in which important parts of the wavefront control loop are tested in different contexts.
A simulation phase covered the prototyping of the wavefront control strategy.
The second phase covers the integration of the actual \gls{rtc} in the simulation environment.
The following phases involve the usage of a telescope simulator at sub-system and system level, see Section~\ref{sec_testing}.
The final integration and test phase will be part of the installation and commissioning at the ELT.

It is intended to test some of the important Soft Real-Time Cluster (\gls{srtc}) functionalities on sky in the context of another instrument, LINC-NIRVANA, at the Large Binocular Telescope (LBT).
Although the \gls{ao} facilities of LBT and ELT have commonalities, they still differ in numerous ways.
The \gls{ao} \gls{rtc} at the LBT is highly integrated into the telescope, and it is neither feasible nor intended to try to test the Hard Real-Time Core (\gls{hrtc}) for \gls{metis} at the LBT.
But the configuration of the \gls{rtc}, the detection of misregistration in the \gls{ao} telemetry, updating the reconstructor in closed loop or the stabilisation of the pupil on the \gls{pwfs} detector are important functionalities that can be tested with LINC-NIRVANA.

\subsubsection{Instrument Testing}\label{sec_testing}

For comprehensive testing of the \gls{scao} sub-system a telescope simulator will be used. This optomechanical setup mimics a number of features of the \gls{elt}, such as the unique pupil shape, a deformable mirror (M4), a field steering mirror (M5), and means of introducing a diffraction-limited light source on the \gls{wfs}.  It provides the complete accessible Field of Fiew (\gls{fov}) that \gls{metis} supplies to the \gls{wfs} (diameter 27\arcsec on the sky).

Different disturbances can be introduced: temporally and spatially variable phase distortions, pupil image motion, and M4 image motion (relative to entrance pupil). For compensation (and, indeed, simultaneously for the introduction of said distortions), an ALPAO 820 deformable mirror and a pupil steering mirror substitute will be used.

\begin{figure} [htb]
   \begin{center}
   \includegraphics[width=0.55\textwidth]
   {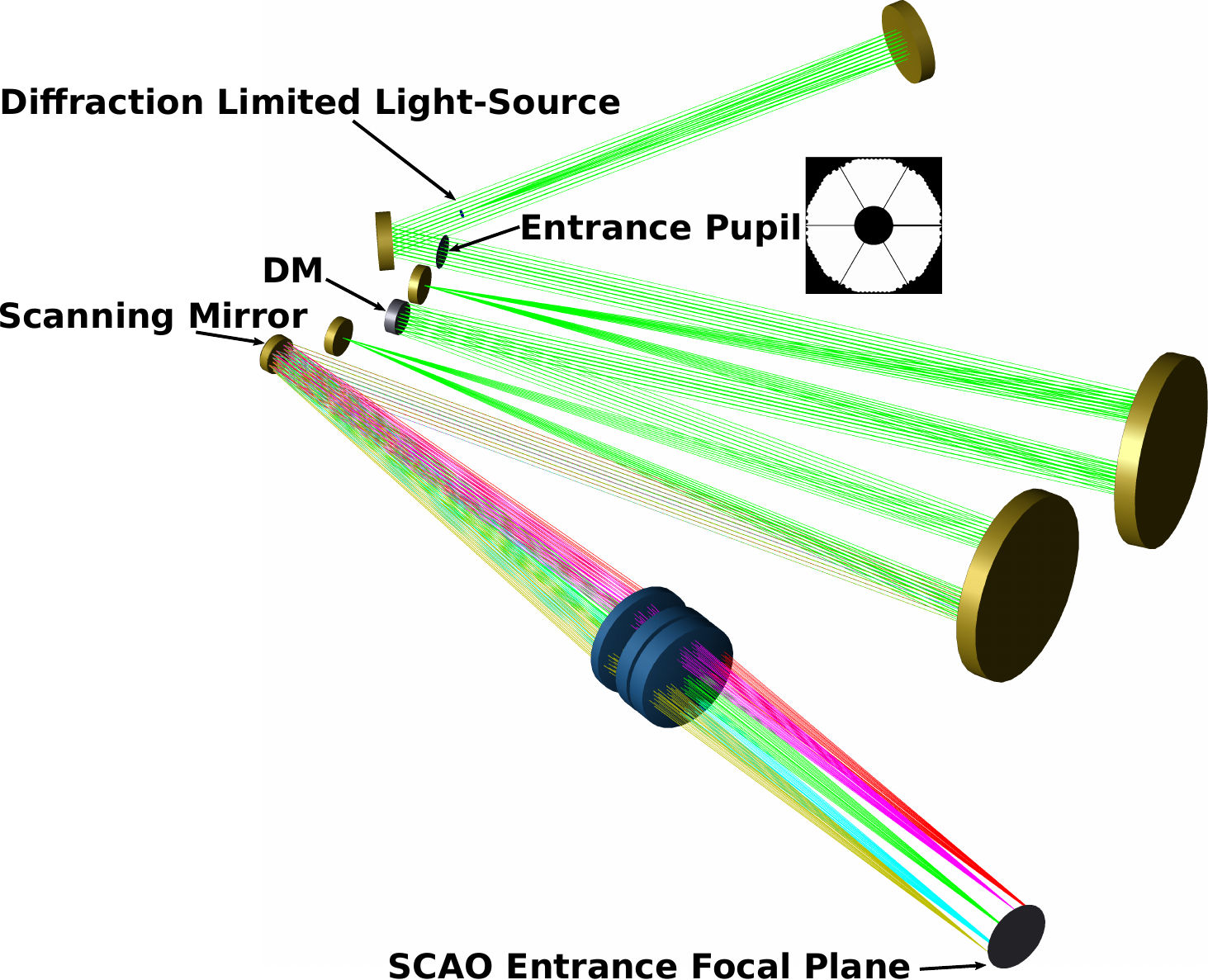}
   \end{center}
   \caption{Optical layout of the telescope simulator. Green: On-axis optical path up to the scanning mirror. The "scanning mirror" doubles as tip-tilt corrector M5,  and together with the lens triplet it provides the complete 27\arcsec FoV required to test SCAO.}
\label{fig:tel_sim_layout}
\end{figure}

Figure~\ref{fig:tel_sim_layout} shows the optical layout of the telescope simulator.
To a large extent, an on-axis-only approach is used, which simplifies the optical design. The final part incorporates a scanning mirror plus a lens triplet that provide the complete accessible \gls{fov} that the \gls{wfs} requires. The optical elements are compact to minimise the overall envelope (1.4 m x 1.4m x 0.5m) and use off-the-shelf optics, opto-mechanics and mechanisms wherever possible. The (mostly) reflective design minimises the chromaticity and enables the simulator to work in the wavelength range between 1.3 and 4~$\mu$m. 

\begin{figure} [htb]
   \begin{center}
   \includegraphics[width=0.55\textwidth]{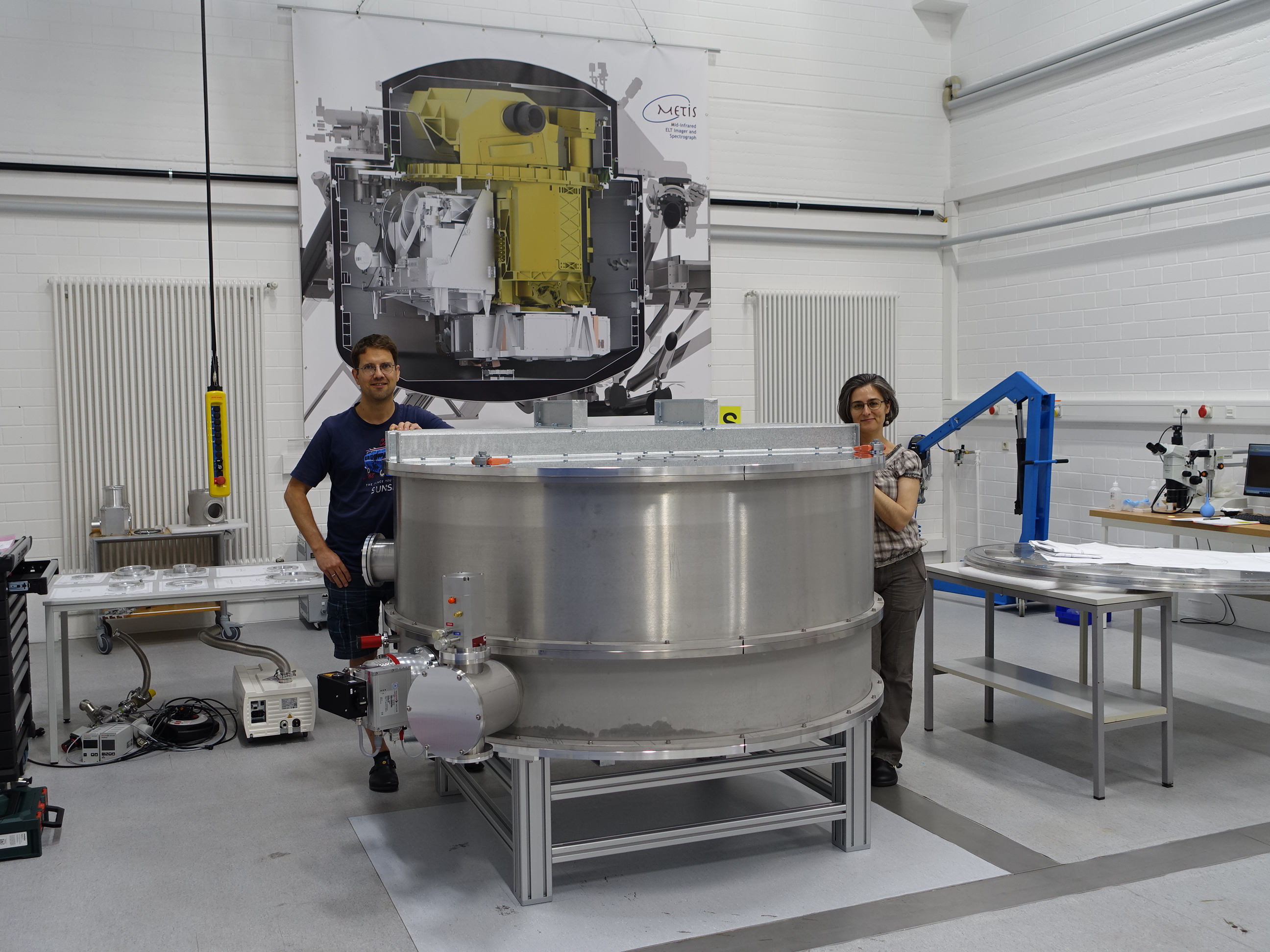}
   \end{center}
   \caption{SCAO test cryostat inside MPIA lab.}
\label{fig:test_cryostat}
\end{figure}

\medskip
The telescope simulator will be used:
\begin{enumerate}
\item As support equipment for \gls{scao} sub-system testing inside the \gls{mpia} lab to
demonstrate \gls{ao} closed-loop operation of the \gls{aocs} and the SCAO Module.
Inside the \gls{mpia} lab the telescope simulator will be in front of a dedicated test cryostat providing operating conditions and hosting the SCAO Module (cf. Figure~\ref{fig:test_cryostat}).
\item As support equipment for system testing to demonstrate the closed-loop operation of the \gls{aocs} and the SCAO Module within \gls{metis}. At the system \gls{ait} facility in Leiden the \gls{wfs} will be inside the \gls{metis} cryostat and the telescope simulator will be in front of it. It will also be used to check the interaction with other subsystems and to test some of the \gls{hci} functionalities.
\end{enumerate}

\section{The SCAO Module Design}\label{sec:scao_module_design}

\subsection{Layout}

The \gls{scao} module is located inside the cryostat of \gls{metis} (cf. Figure~\ref{fig:metis_fdr_design}). A cold dichroic AO pick-off mirror
immediately in front of the \gls{scao} module is used to separate the near-infrared part of the light,
which is used for wavefront sensing. The \gls{scao} module provides a \gls{pwfs} as well as opto-mechanical actuators for field selection and modulation of the \gls{ngs} in the field of view.


\subsection{Component Overview}

\begin{figure}[h!]
\begin{center}
\includegraphics[width=0.65\textwidth]{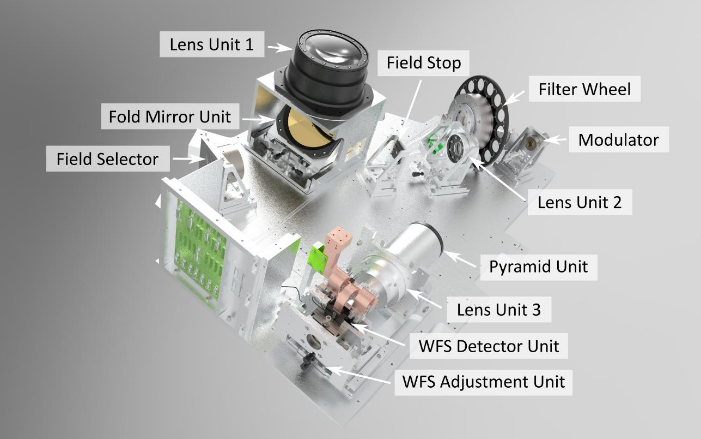}
\caption[SCAO Module FDR design]{SCAO Module FDR design}
\label{fig:ScaoModuleFDRDesign}
\end{center}
\end{figure}

The \gls{scao} module subsumes the physical components that are required in the instrument domain to provide \gls{scao} wavefront correction for \gls{metis}.
An overview of the units in the \gls{scao} module is shown in Figure \ref{fig:ScaoModuleFDRDesign}. 

\begin{description}
\item[Base Assembly:] The main structure to which all other units are mounted to. It also provides the mechanical interface to METIS' common cold structure, the housing of the \gls{scao} module, the \gls{scao} module internal harnesses and infrastructure for temperature sensing and heat flow.
\item[Lens Unit 1:] A lens group at the entrance of the \gls{scao} module which re-images the Common Fore-Optics (\gls{cfo}) Focal Plane 1 (\gls{fp}1) to the SCAO subsystem (SCA) \gls{fp}1 and which forms a conjugate pupil of the \gls{elt} entrance pupil at \gls{sca} Pupil Plane 1 (\gls{pp}1).
\item[Fold Mirror Unit:] A flat mirror introduced to fit the optical train of the \gls{scao} module into a more compact envelope.
\item[Field Selector:] A controllable tip-tilt mirror in the \gls{pp}1, used to center the \gls{ngs} on the \gls{pwfs}.  The field selector allows to acquire the \gls{ngs} anywhere in the \gls{fov} that is provided by the \gls{cfo} in \gls{sca} \gls{fp}1.
\item[Field Stop:] A mask in the \gls{sca} \gls{fp}1 that limits the field that is propagated towards the \gls{pwfs}, mainly to minimize stray light. 
\item[Lens Unit 2:] A lens group which reimages \gls{sca} \gls{fp}1 to  \gls{sca} \gls{fp}2, the tip of the pyramid with an f-number of 70. It also forms a conjugate pupil of the \gls{elt} entrance pupil at the \gls{sca} \gls{pp}2.
\item[Filter Wheel:] The wheel allows to introduce different band pass and neutral density filter combinations for the wavefront sensing wave band.
\item[Modulator:] A fast tip-tilt steering mirror used to modulate the position of the \gls{ngs} Point Spread Function (\gls{psf}) on the tip of the pyramid. The modulation radius allows to balance the linear range of the \gls{wfs} and its sensitivity. The modulator is located in \gls{sca} \gls{pp}2.
\item[Pyramid Unit:] Provides the core component of the \gls{pwfs}. The four faceted pyramid is located in \gls{sca} \gls{fp}2 and directs the light of the incoming beam into different directions, depending on the local instantaneous phase distribution.
\item[Lens Unit 3:] A lens group which re-images the ELT primary mirror (\gls{m1}) to the \gls{wfs} detector.
\item[WFS Detector Unit:] The SAPHIRA 320$\times$ 256 pixel array in its housing and the required components for temperature control.
\item[WFS Adjustment Unit:] A mechanical structure that holds the \gls{wfs} detector unit. It provides means to manually adjust the position of the detector and the entire \gls{wfs} during the alignment.
\end{description}

Some of the components are explained in a bit more detail below:

\subsubsection{Pyramid}

\begin{figure} [htb]
   \begin{center}
   \includegraphics[width=0.45\textwidth]{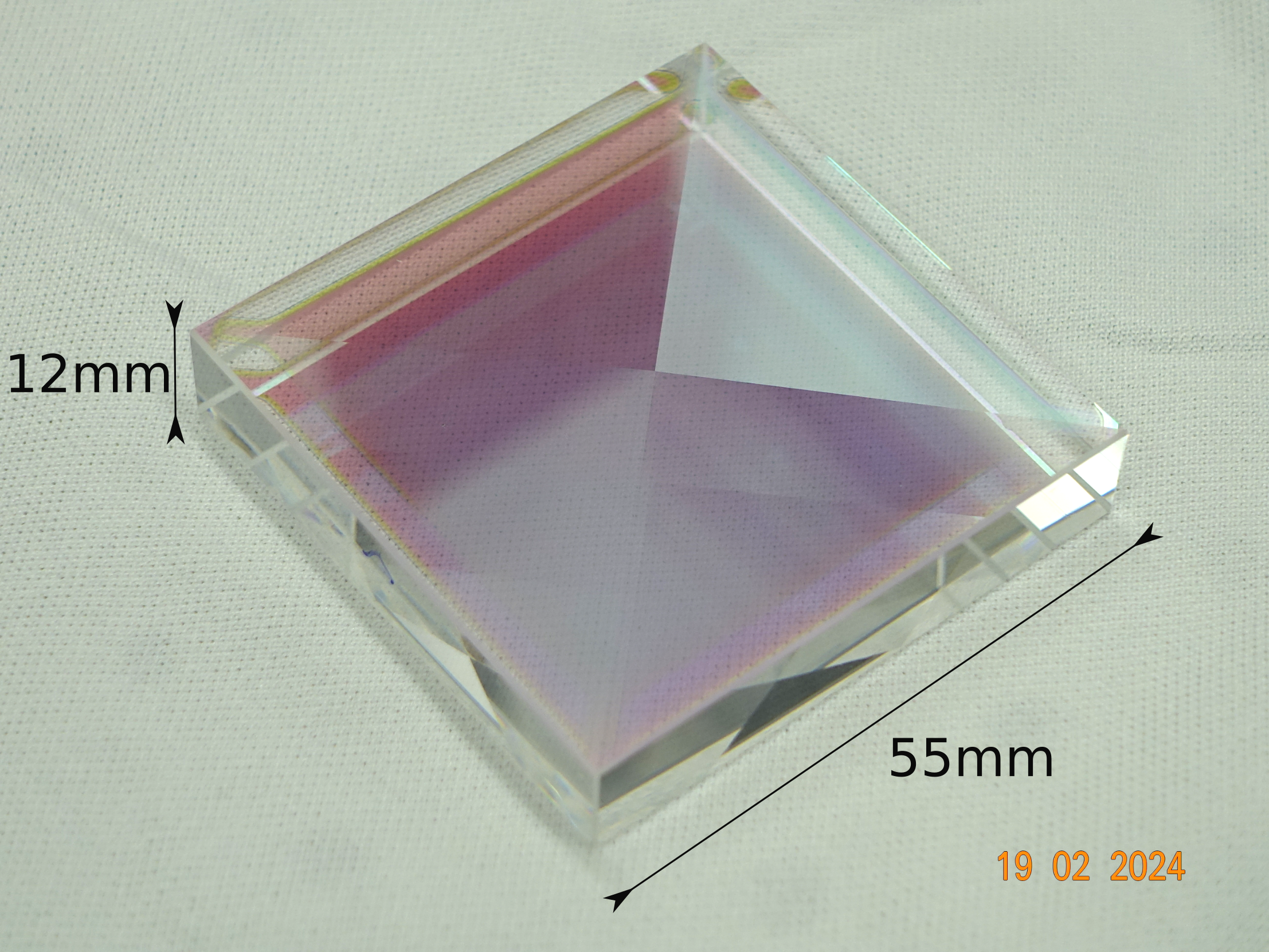}
   \end{center}
   \caption{The pyramid prism.}
    \label{fig:pyramid}
\end{figure}

The Pyramid Unit contains the core component of the \gls{pwfs}:  a four-sided pyramid prism with demanding requirements (cf. Table~\ref{tab:PyramidSpecs}). It works in transmission -- the four facets deflect the incoming beam in four different directions. The separation of the light in the transitions across neighbouring facets decomposes the shape of the wavefront in two orthogonal directions. The subsequent Lens Unit 3 creates the four replicas of the pupil which carry the information of the wavefront deformation. 

The actual pyramid prism is shown in Fig.~\ref{fig:pyramid}.

\begin{table}[ht]
  \footnotesize
  \caption[Pyramid prism basic specification.]{Pyramid prism basic specification.}
  \label{tab:PyramidSpecs}
  \centering
  \begin{tabular}{>{\raggedright\arraybackslash} p{4cm}l}
  \toprule  
  \textbf{Parameter}        & \textbf{Specification}\\
  \midrule
  Pyramid angle $\alpha_P$           & 1.893$\deg \pm$ 5 arcsec \\
  ($\alpha_1 \approx \alpha_2 \approx \alpha_3 \approx \alpha_4$) & $\le$ 10 arcsec\\
  Orthogonality angle $\Delta\phi$ & $\le$ 5 arcsec \\
  Pyramid angle centering &  $\pm$ 0.1 mm \\
  Edge sharpness $\epsilon$, size  &   $\leq$ 10 µm, no chips \\
  (edge between pyramid faces)\\
  Roof sharpness, tip size       & $\leq$ 10 µm \\
  \bottomrule
 \end{tabular}
\end{table}

\subsubsection{Filter Wheel}

The filter wheel allows us to select different band pass and neutral density filter combinations for the wavefront sensing waveband.
The standard wavefront sensing band is the $K$-band. 
To be able to observe also very bright stars, a neutral density filter is needed to avoid detector saturation. In the opposite case, with wavefront sensing at the faint end with reduced performance, a combined $H+K$ filter can increase the limiting magnitude.

The Filter Wheel is a unit which is based on the Indexed Cryogenic Actuator for Rotation (\gls{icar}) \citep{barriere13}, the \gls{metis} standard mechanism for wheels. 
There will be four filters: $H$-band, $K$-band, $H$+$K$ band, and a neutral density filter with an average transmission of 6.3\% across the $K$-band.

\subsubsection{Field Selector}

The field selector is a controllable tip-tilt mirror in SCA \gls{pp}1, which allows to center any point in the accessible \gls{fov} (diameter 27\arcsec on the sky) on the \gls{wfs}.

Table~\ref{tab:FSSpecs}  lists the main specifications for the field selector actuator. 

\begin{figure} [htb]
   \begin{center}
   \includegraphics[angle=-90,width=0.45\textwidth]{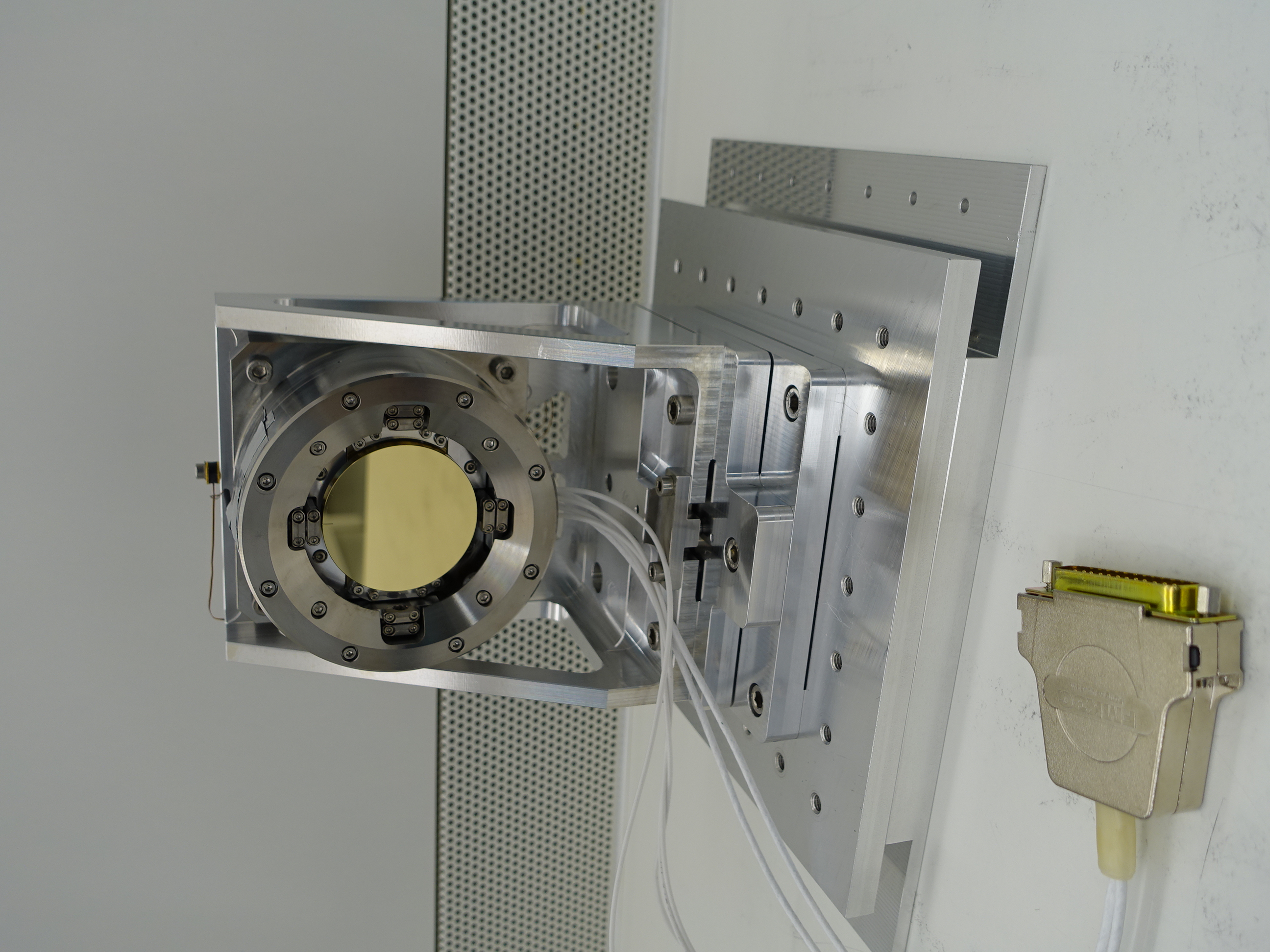}
   \end{center}
   \caption{Field Selector prototype.}
\label{fig:FieldSelector}
\end{figure}

\begin{table}[ht]
\caption{Field Selector hardware specifications}
\label{tab:FSSpecs}
 \begin{tabular*}{0.9\columnwidth}{ll}
 \toprule
  \emph{steering platform}             & tip-tilt, orthogonal \\
  \emph{range ($\theta_x$,$\theta_y$)} & 0°-6°                \\
  \emph{repeatability}                 & $<$ 3.3~µrad (PV)      \\
  \emph{position stability}            & $<$ 0.67~µrad (PV)     \\
  \emph{max. velocity}                 & $>$ 7~mrad/s           \\
  \bottomrule
 \end{tabular*}
\end{table}
The  core component of the field selector is the tip-tilt mechanism, a gimbal mounted mirror platform, supported by C-flex bearings, which positions the field selector Mirror. The main body of the mechanism is made from titanium
alloy, the mirror, the C-flex bearings, and the mirror platform are made from aluminium alloy. The aforementioned gimbal also serves to decouple thermal stress between the mirror and the tip-tilt platform of the mechanism.
Figure~\ref{fig:FieldSelector} shows the prototype mechanism produced by the company Physik Instrumente.

The challenge of this application is the combination of a large travel range in combination with a high angular resolution operating in a cryogenic environment.

As internal measurement system for the field selector actuator a KD-5100 Eddy current system from Kaman is implemented.

An angular drive is implemented that moves the pivoting platform with the help of Piezo walk drives (PICMAWalk drive N-331 by Physik Instrumente).   

The usage of Piezo Walk Drives, where a number of piezo stacks are commanded in a coordinated fashion to move the actual actuator, requires a special controller that supports this type of drive. The controller is also provided by Physik Instrumente.

\subsubsection{Modulator}

The modulator is a controllable tip-tilt mirror in SCA-\gls{pp}2.
Its task is to modulate the light of the star over the four facets of the pyramid.
A sinusoidal tip-tilt motion of the mirror will cause the  image of the \gls{ngs} to move along a circular path around the tip of the pyramid. The maximum modulation radius required for the Modulator is 10$\lambda$/D (which translates into 0\farcs123 on-sky radius at the WFS' maximum operating wavelength).

Table~\ref{tab:ModSpec} lists the main specifications derived for the tip-tilt mechanism.

\begin{table}[ht]
\caption{Modulator hardware specification. }
\label{tab:ModSpec}
 \begin{tabular}{ll}
 \toprule
  \emph{steering platform}             & tip-tilt, orthogonal \\
  \emph{range ($\theta_x$,$\theta_y$)} & 0 mrad - 3 mrad      \\
  \emph{repeatability}                 & 2~µrad (PV)          \\
  \emph{minimum incremental motion}    & $<$ 1~µrad             \\
  \\
  \emph{motion}                        & continuous circular  \\
  \emph{operating frequency$^1$}           & 100~Hz -- 1000~Hz    \\
  \bottomrule
 \end{tabular}
\footnotetext[1]{Full circles per second!}

\end{table}
The core component of the modulator is the tip-tilt actuator which positions the modulator mirror. 
The modulator actuator is based on an existing product series, the S-330 tip-tilt steering mirror platform series from Physik Instrumente. 
Two piezo actuators per axis are working in push-pull mode supporting the mirror platform and provide a high angular stiffness -- which is required for the high operating frequency of this application.
The actuator is made from titanium alloy, the mirror from aluminium alloy. 
A single bolted connection in the center of the mirror avoids issues with the different \glspl{cte} of the used materials.

A crystalline piezo material is used for this cryogenic application. The material has a number of advantages over conventional piezo ceramics:
\begin{itemize}
 \item A reduced dependency of the piezo amplitude on temperature, which restricts the loss of travel range in the cryogenic operating environment.
 \item Small dielectric losses inside the piezo material, resulting in low power consumption and very low heat dissipation, which is important especially at high operating frequencies.
 \item The response of the piezo material is highly linear and almost hysteresis free -- allowing for open-loop scanning applications.
\end{itemize}

\begin{figure} [htb]
   \begin{center}
   \includegraphics[width=0.45\textwidth]{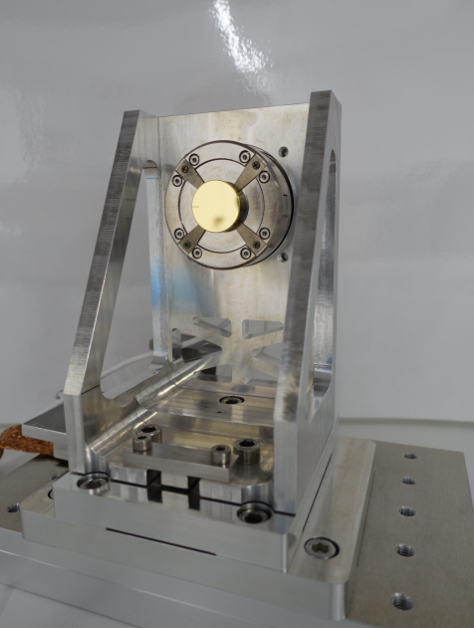}
   \end{center}
   \caption{Modulator prototype.}
    \label{fig:Modulator}
\end{figure}

Figure~\ref{fig:Modulator} shows the modulator actuator on its support. 
The design of the support provides the same alignment features as the support for the field selector. Its position on the base plate is defined by the corresponding reference dowel pin in the base plate. It constraints the location of the mirror surface along the folded optical path. 
The range of the actuator is much smaller than the range of the field selector. 
This must be taken into account in the alignment of the \gls{scao} module. 

    \section{Wavefront Control Strategy \label{sec:wfc-strategy}}



As with any other \gls{ao} system installed on the \gls{elt}, the \gls{metis} control strategy contemplates the main control loop and a set of auxiliary supporting loops \citep{correia22}.

When devising our strategy we had the following goals in mind:
\begin{enumerate}
\item Controlling the maximum number of \gls{m4} degrees-of-freedom up to 5352
without generating unwanted spurious signals (e.g. differential-piston
modes)
\item Performing numerical derotation and streamlined on-the-fly reconstructor
updates, including management of mis-registrations between M4 and the \gls{wfs}.
\item Using a set of aperture-spanning modes as a control basis ranked jointly
by increasing spatial frequency and force impinged on \gls{m4}
\end{enumerate}

Although operating the \gls{pwfs} in the $K$-band makes wavefront sensing / reconstruction easier in many respects, notably by providing the ability to avoid pupil fragmentation modes (differential piston or petal modes, see, e.g. \cite{bonnet18}), meeting jointly all the goals listed above required the development of a custom, near-optimal (in the minimum residual variance sense) solution. Such aspects are covered next.

\subsection{Main Control Loop}\label{sec:Main_Control_Loop}

\begin{figure}[htpb]
	\begin{center}
            \includegraphics[width=0.5\textwidth]{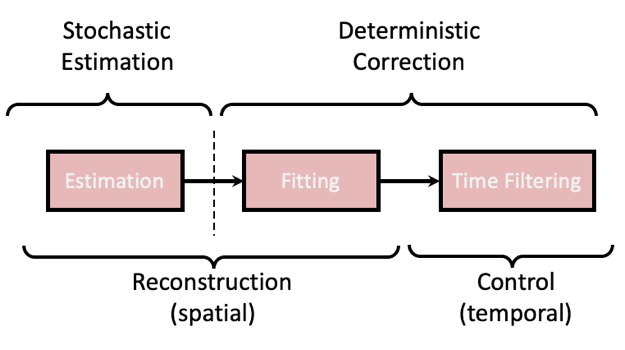}
	\end{center}
	\caption[Wavefront control operations]
	{\label{fig:conceptual_split_estimation_control}
     Separation between stochastic estimation and deterministic correction, spatial reconstruction and temporal, dynamic control.   This
simplification shows the interplay of the different, spatial
and temporal, stochastic and deterministic operations. The estimation
and fitting, whereas both spatial filtering operations, are two
distinct steps. This is a feature we retained since the
pupil and DM actuator mesh optically rotate in front of the
P-WFS. Decoupling the reconstruction from the fitting is an important aspect that is considered in our developments, as it allows a streamlined implementation and AO system adaptation to changing
     conditions.}
 \end{figure}

To meet all control goals established for \gls{metis}, we have adopted a tailored implementation of the classical \textit{minimum mean square error} (MMSE) wavefront reconstruction strategy \citep{ellerbroek09, correia22}. In \gls{metis} we make explicit use of disjoint vector spaces for the (static) wavefront reconstruction from \gls{pwfs} "slopes" and the \gls{dm}-fitted control modes. In more detail, \gls{metis} uses a set of spatially compact, zonal, bi-linear spline influence functions (orthonormal) centred at the crossings of a regular, Cartesian grid aligned with the pyramid sensor sub-apertures (a solution very similar to \cite{ellerbroek02} featuring also a squared Laplacian regularisation matrix) followed by a (once again) regularised projection on the \gls{m4} actuator spanned space \citep{obereder23} via a set of force-minimising Karhunen-Loève control modes \citep{verinaud23}. Finally, temporal filtering, which incorporates leaky proportional-integral regulation, is applied to residual modal expansion coefficients. Fig. \ref{fig:conceptual_split_estimation_control} schematically represents these steps. Detailed information can be found in \cite{correia22}. 

A particularity is that \gls{metis} will numerically derotate the pupil and, in passing, compensate likewise for the remaining misregistrations such as lateral offsets, absolute and differential magnification and eventually higher-order anamorphisms if necessary \citep{2021MNRAS.504.4274H}. In order to do just that, the reconstruction step is further subdivided into two steps: i) the
stochastic wavefront estimation from \gls{pwfs} data on a Fried layout, bilinear-spline influence function Virtual Deformable Mirror (\gls{vdm}), which is independent of \gls{m4}, and ii) the
 fitting/projection onto the desired control space. The latter takes into account the relative position of \gls{m4} with respect to the fixed \gls{vdm} reconstruction space, making for a very straightforward implementation. The same regularisation principle that is used for the wavefront reconstruction (i.e. utilize the know statistics of the wavefront) is used in this projection step.

In compact format, the \gls{metis} \gls{rtc} hard real-time pipeline will implement the reconstruction step by computing the \gls{vdm} command vector $\uvec_k^{\VDM}$ and the error signal $\evec_{k}$ at time step $k$ as

\begin{equation} \label{eq1}
\begin{split}
\uvec_k^{\VDM} &  = \left( \D_{\VDM}^\T  \D_{\VDM} + \alpha_{\mathrm{rec}} \Delta_{\VDM}^{2} \right)^{-1} \D_{\VDM}^\T \svec_k \\
\evec_{k}  & =  \MtoC_0^\dag \left( \N_{\Mfour}^\T  \N_{\Mfour} + \alpha_{\mathrm{proj}} \Delta_{\mathrm{M4}}^{2} \right)^{-1} \N_{\Mfour}^\T \N_{\VDM}(\mathcal{P})\uvec_k^{\VDM} 
\end{split}
\end{equation}
 where $\D_{\VDM}$ is the \gls{pwfs} forward model, $\alpha_{\mathrm{rec}} \Delta_{\VDM}^{2}$ 
 and $\alpha_{\mathrm{proj}} \Delta_{\mathrm{M4}}^{2}$ are discrete $\alpha$-weighted bi-harmonic regularisation matrices \citep{ellerbroek02} that operate on the \gls{vdm} and M4, respectively. $\MtoC_0$ is the control modes to \gls{m4} influence function matrix computed in \cite{verinaud23}, $\svec_k$ are the \gls{pwfs} slopes-maps \citep{fauvarque16} and $\N_{\Mfour}$ and $\N_{\VDM}$ are respectively the concatenation of \gls{m4} and \gls{vdm} influence functions. The latter projection compensates numerically for misregistration, decomposed as a set of parameters $\mathcal{P} =  \{\Delta_x,\Delta_y,\theta, \cdots\}$.

 Time filtering then ensues, with Proportional-Integral (PI) control for the Tip-Tilt (\gls{tt}) and High-Order (\gls{ho}) modes. Anti-windup is ensured through error deflation using the CCS feedback
   \begin{equation}\label{eq:IIR_filter}
   \mvec_k = \sum_{j=0}^{P_{IIF}} b_j \evec_{k-j} - \sum_{i=1}^{Q_{IIF}}a_i \mvec_{k-i},
     \end{equation}
     where $\evec_{k-j}$ is the error signal expressed in the control
   modal basis after ($\mu_{\text{aw}}$-weighted) error deflation is applied 
   \begin{equation}\label{eq:error_with_antiwindup}
   \evec_k = \evec_k^ {-} - \mu_ {\text{aw}} \boldsymbol{\epsilon}_{k}
 \end{equation}
 The upper-script $^-$ expresses the error before the
   windup correction is applied and $\mu_ {\text{aw}}$ is a weighting applied to 
   \begin{equation}
 \boldsymbol{\epsilon}_{k}= {\mvec}_{k-1} - \widehat{\mvec}_{k-1}
 \end{equation}
 where:
 \begin{itemize}
 \item ${\mvec}_{k-1}$ are \gls{metis} {\it requests\/} expressed in the  control
 modal basis. We call them requests to account for the subtle, yet real, difference between what is requested and what ends up being actually applied past \gls{ccs} \gls{m4} saturation handling.
 \item $\widehat{\mvec}_{k-1}$ is the “echo” signal sent back by the
 \gls{ccs} corresponding to the actually applied commands, expressed in the
 instrument control basis. These may differ from the \textit{requests} on account
 of exceeding operational limits  as per \cite{ESO-311982-V3};
 \item $\mu_ {\text{aw}}$ is a scalar weight whose default value is 1
 \end{itemize}

The deployment of predictive control (either in the form of the
optimal Minimum-Variance solution or otherwise) is not our baseline
and its investigation is deferred to a later stage. 



\subsection{Auxiliary Control Loops \label{sec:auxiliary-loops}}
The auxiliary control loops refer to supporting loops that stabilise
or adjust working setpoints needed for optimal performance. 
Due to the nature of the
auxiliary control loops that involve system telemetry processing and/or
control actions on actual hardware, they will be detailed
separately below. Software optimisation tasks are covered in \S \ref{sec:SRTC}.

\subsubsection{Pupil Position Control Loop} \label{sec:PPC}
The Pupil Position Control (\gls{ppc}) drives the Pupil Stabilisation Mirror (\gls{psm}) to stabilise the pupil with a signal derived from
processing the \gls{wfs} pixel intensities.

Our baseline prototyped and tested in \cite{coppejans23} consists in using matched filters in a similar
implementation to that of TMT \cite{veran2017telescope}. Yet, in our case the template
filters fully rely on the (jagged) inner and outer edges of the primary
pupil with removed spiders, which proved more robust after substantial analysis reported in \cite{coppejans23}. 



\subsubsection{Differential Tip-Tilt Control Loop}\label{sec:DiffTTLoop}
Any differential tip-tilt between science and \gls{wfs} path is compensated for when using any of the \gls{hci} modes.
\gls{qacits} \citep{huby15} will generate a TT signal which is added to the modulation signal
after time filtering. The mathematical description being so elementary, we allow ourselves to omit it here.  
Compensating measured tip/tilt terms by offsetting the modulation centre avoids the use of \gls{wfs} signal offsets, which would otherwise lead to a reduction in the \gls{wfs} dynamic range.


\subsubsection{Differential High-Order Control Loop - NCPA correction}\label{sec:DiffHOLoop}
The differential high-order control loop compensates for \gls{ncpa} 
and water-vapour seeing via Pyramid Wave Front Sensor (\gls{pwfs}) slope offsets. It is to be
deployed in all HCI-modes and normal IMG if requested. 

The \gls{alwfs} algorithm \citep{Orban+2024} is used to compute the \gls{pwfs} reference signal $\svec_{\Delta}$, assuming a nominal optical gain of 1 by
projecting up to 100 Zernike modes onto the set of slopes \gls{pwfs}.
\begin{equation}
  \svec_{\Delta} = \D \N_{VDM}^\dag \Psi_{Zer}^\T \vec{a}
\end{equation}
where
\begin{itemize}
\item $\Psi_{Zer}$ is a concatenation of Zernike modes expressed in
phase space at the nominal pupil resolution
\item $\vec{a}$ is the Zernike mode coefficient set resulting from the
projection of the NCPA or water-vapour optical aberration onto the set
of Zernike modes
\end{itemize}

Note that compensating for NCPA means that the SCAO will see a significant amount of aberration at the tip of the Pyramid WFS. This is expected to reduce the SCAO performance, and thereby increase the rms of residual phase errors.  This is quantified in Sec.~\ref{sec:sim-ncpa}.

%





\subsection{Transients \label{sec:transients}}

A number of transient phenomena will occur where the closed-loop control will transition from one state to another in a short period of time. Most notably this is the case when the control of the telescope and it's wavefront correcting elements is taken over from internal control by the ELT's guide probe. Another very important transient is the so-called Recurrent Optimisation of Unit Stroke (ROUS) event, where the adaptive elements are periodically re-set close to their center positions after the closed-loop operation may have driven them close to the edge of their respective actor ranges, all while maintaining the currently demanded wavefront shape.  What are not regarded as transients are events triggered by our internal control system auxiliary control loops (see Sec.~\ref{sec:auxiliary-loops}) or the loop co-processing tasks (see Sec.~\ref{sec:loop_coprocessing_tasks}).

\subsubsection{Telescope Handover}


\gls{metis} plans to hand over from the telescope following a sequence of steps as outlined below:
\begin{enumerate}
\item The \gls{ppc} loop is engaged with the pupil stabilised on the
\gls{pwfs} detector 
\item Handover control of a set of low-order modes including tip-tilt which add to a reference frozen by the telescope just before handover (typically within a few frames delay)
\item Trigger the mis-registration algorithm (cf. \S \ref{sec:loop_coprocessing_tasks})
\item Update reconstructor and start controlling a larger number of modes (expected a factor of 10-20x the number of modes). 
\item Repeat 4. and 5. until the pre-defined number of control modes
is achieved
\end{enumerate}

During this handover phase, we likely need to update the control matrix
based on the mis-reg parameters estimated during loop closure. This
may result from any differential motion and/or optical warping between
the pupil and M4, which will then be estimated by our adaptation of SPRINT \citep{2021MNRAS.504.4274H}.
Otherwise the pupil stabilised in lateral motion by the \gls{ppc} loop
will grant only a sub-pixel residual. The observation
model from which the rotation is inferred will then be used to
numerically derotate the pupil via our DM projection step.

We note that any unknown lateral motion and/or magnification will have
no implication on our interaction matrix calibration strategy (see Sec.~\ref{sec:SRTC}). The latter depends on the diffraction
pattern generated by the pupil \gls{m1} whose transverse translations are stabilised with the
\gls{ppc} loop. Any mis-registrations will then be dealt with
independently in the projection step.
\subsubsection{ROUS}

The baseline strategy consists of considering the Recurrent Optimisation of Unit Stroke (\gls{rous}) transient
in much the same way that we consider the bootstrapping transient after
telescope handover. All tests performed leading to \gls{fdr} were successful in that the control parameter choices were shown to be robust to handle \gls{rous} without any need to control fewer modes, let alone opening the loop and giving control back to the telescope. Nevertheless, we are ready for such strategies should they become necessary due to operational constraints. 

\paragraph{Scheduling the ROUS}
\gls{metis} will schedule observations \textit{around} \gls{rous} events such
that they are not affected by such events. In order to enforce it, \gls{metis} will influence
the time of execution to avoid negative impact on the ongoing
observations (as per \cite{ESO-311982-V3}).

\paragraph{Control loop adaptation}
During the \gls{rous}, \gls{metis}' control loops will remain
closed. Although our simulations show that the loop remains stable
across the event with all modes controlled, in practice the
\gls{rous} will be crossed with fewer actively controlled
modes or a combination of strategies similar to the handover
transient. 

In the likely event that the pupil and the misregistration parameters
change on account of the optical offloading, we will first stabilise the pupil to its
nominal position, then deploy the misregistration identification
algorithm with the loop closed on a small subset of modes. 

The transient lasts around 2-3\,s during which we expect a tilt
excursion of $<$ 10 mas. Once this transient is over, pupil stabilisation and
mis-registration management (if needed)  are carried out. Past
this period, nominal operations can resume.

\subsection{Loop co-processing tasks} \label{sec:loop_coprocessing_tasks}
The \gls{metis} wavefront control strategy is supported by a number of \textit{Loop
Co-Processing} monitoring and optimisation tasks. These allow
operating the system within requirements and allocated tolerances. In addition to
the \textbf{Auxiliary Control Loops}, tasks under the following
categories will be executed in
parallel to the control loops:
\begin{enumerate}
\item\textbf{System parameter estimation} -- encompassing the tasks where
  geometric and optical parameters are estimated from telemetry and
  metrology accessible in the system, in particular the registration between M4 and the \gls{wfs} grid which will be determined by injecting test signals into the closed-loop correction, similarly to the SPRINT method described in \cite{heritier21}.
  \item \textbf{Control loop parameter optimisation }-- covering the tasks
    pertaining to the optimisation of control matrices, loop gains etc
    \item \textbf{Statistics $\&$ Diagnostics} -- where performance metrics are made
  available from the AO telemetry, science detectors and other
    internal metrology.
\end{enumerate}

A more in-depth list of such tasks is provided in Table \ref{fig:srtc-table}
found further below in Sec.~\ref{sec:SRTC}.


\section{Real-Time Computer \label{sec:rtc}}
The \gls{rtc} was developed not only with the goal of meeting the defined performance requirements, but also to have a solution with the lowest possible complexity for maintainability reasons. As a consequence, we use exclusively commercial off-the-shelf (COTS) hardware and rely on standard third-party software libraries to arrive at the tailor-made solution we describe in this section.

\subsection{System Context of the RTC}

\begin{figure}[ht]
  \begin{center}
  \includegraphics[width=0.7\columnwidth,
                   clip,
                   trim= 0.0cm 1.7cm 0cm 0.5cm]{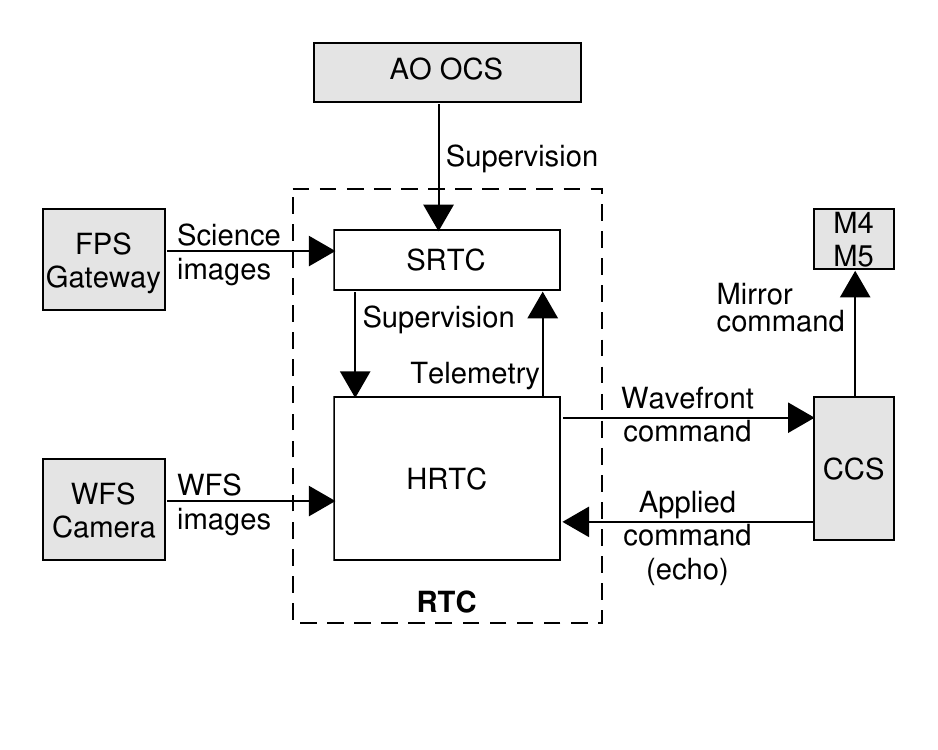}
  \end{center}
  \caption[METIS RTC context] 
    {\label{fig:RTC_Context}
METIS RTC system context. }
\end{figure}

The RTC consists of two parts: the \gls{hrtc} and the \gls{srtc}.
While the \gls{hrtc} runs the main AO loop with the tight timing constraints described in Section \ref{sec:hrtc_problem_size}, the \gls{srtc} supervises and optimises the \gls{hrtc} operation.
The environment of the \gls{rtc} and its principal data flows are depicted in figure~\ref{fig:RTC_Context}.

In the main AO data path,  the \gls{wfs} camera transmits \gls{wfs} images from which the \gls{hrtc} computes wavefront commands for the \gls{ccs}.
The \gls{ccs} checks the wavefront command for safety and 
responds with the applied commands to the \gls{hrtc}.
Furthermore,  the Focal Plane Sensor GateWay (\gls{fpsgw}) provides science images to the \gls{rtc} in order to enable the focal plane wavefront sensing mentioned in Sec.~\ref{sec:design_decisions} via the auxiliary control loops described in Sec.~\ref{sec:auxiliary-loops}.
At the top, the AO Observation Coordination System (\gls{ao ocs}) supervises the \gls{rtc} by coordinating, monitoring and configuring the \gls{srtc}.

\subsection{HRTC}

\subsubsection{Problem Size}
\label{sec:hrtc_problem_size}

\acrshort{metis} \acrshort{scao} uses a \acrshort{pwfs} running at 1\,kHz. Each WFS frame has a size of 192\,x\,192 pixels as int32, where each pupil has 6,376 subapertures and a diameter of 90 subapertures. 
The HRTC runs the main AO loop as described in Section~\ref{sec:Main_Control_Loop}.
As a result,  it produces a \acrshort{ccs} wavefront command vector with 4,866 elements, including tip and tilt, within a maximum RTC computation time of 909\,\textmu{}s, i.e.
from last first WFS data received by the RTC and the last command data sent out by the RTC.
The size of the Command Matrix (\gls{cm}) is 4,866 x 12,752, roughly 249\,MB as float32.

Based on the these requirements, the performance requirements for the HRTC is 137\,GFLOP/s, with a memory transfer throughput of 276\,GB/s. 
The operational intensity~\citep{ApplyingTheRooflineModel2014} of the Wave Front Control (\gls{wfc}) task is 0.497. 
That means that the \gls{wfc} computing demand is dominated by the Matrix Vector Multiplication (\gls{mvm}) of the command matrix and the \gls{wfs} signal vector.



\subsubsection{Prototype Hardware}
\label{sec:hrtc_hardware}

The \gls{hrtc} hardware consists of an Asus Barebone ESC4000A-E10 system, with an AMD EPYC 7542 processor and 512 GB of DDR4-3200 memory. Two Nvidia A100 (40GB) Graphical Processing Units (GPUs) are used to accelerate the \gls{mvm} of the \gls{cm}.

\subsubsection{Prototype Software}

The HRTC main software is written in C++. 
It is designed as a pipeline as depicted in
Figure~\ref{fig:HRTC_WFC_computing_pipeline} and consists of the following main components:

\begin{figure}[ht]
  \begin{center}
    \includegraphics[width=0.7\columnwidth,
                     clip,
                     trim= 0cm 1.5cm 0cm 0.5cm]{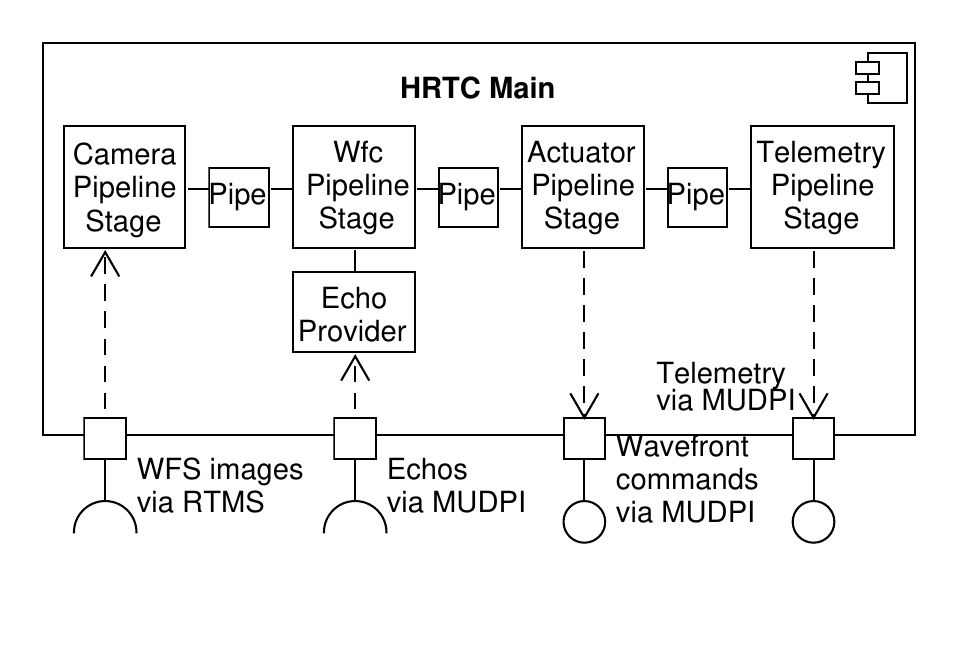}
   \end{center}
   \caption[HRTC WFC computing pipeline]
     {\label{fig:HRTC_WFC_computing_pipeline}
                HRTC WFC computing pipeline structure.}
\end{figure}

\noindent
\begin{itemize}
    \item \texttt{CameraPipelineStage} receives the \gls{wfs} images via the Real-Time MUDPI Stream Protocol (\gls{rtms}). 
    
    \item \texttt{WfcPipelineStage} computes the wavefront commands from the WFS images using Nvidia's Compute Unified Device Architecture (\gls{cuda}) and Advanced vector Extensions 2 (\gls{avx2}). 
    \item \texttt{ActuatorPipelineStage} sends out the wavefront commands using the network protocol \acrshort{mudpi}.
    \item \texttt{TelemetryPipelineStage} sends out the telemetry records via a Multicast UDP Interface (\gls{mudpi}). 
    \item \texttt{CommandEchoProvider} receives the \acrshort{ccs} echo stream via \gls{mudpi} and passes it to the \texttt{WfcPipelineStage}. 
    \item \texttt{Pipe} stores a configurable number of elements. 
\end{itemize}


\subsubsection{Real-Time Performance Check}
The goal of the real-time performance check is to verify if the HRTC meets its performance requirements described in section~\ref{sec:hrtc_problem_size}.  

\begin{figure}[ht]
  \begin{center}
    \includegraphics[width=0.6\columnwidth,
                     clip,
                     trim= 0cm 0.5cm 0cm 0.4cm]{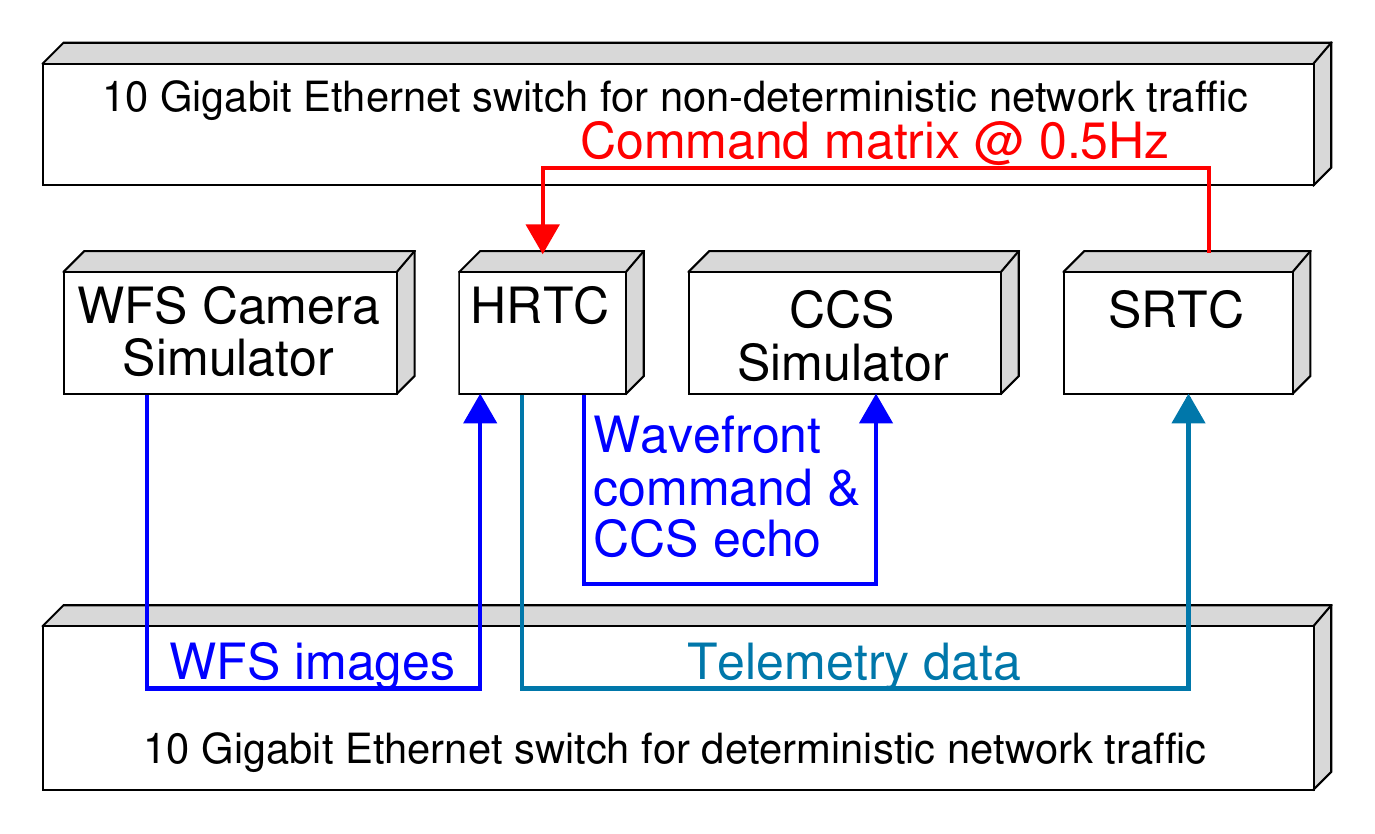}
  \end{center}
  \caption[Principal data flows during the \gls{hrtc} performance test]
  {\label{fig:performance_test_setup}
Principal data flows during the \gls{hrtc} performance test.
}
\end{figure}

For this,  the \gls{hrtc} is put into a computer network that resembles 
a possible setup at the \gls{elt}.
The network layout and the data flows are depicted in Figure~\ref{fig:performance_test_setup}.

The experiment lasted for five minutes in which the \gls{wfs} simulator generated 300,000\,images. 
The most important real-time performance requirement of the \gls{metis} \gls{hrtc} is the RTC computation time for the \gls{wfc} loop. It must be below 909\,\textmu{}s. The current \gls{hrtc} measures the RTC computation time by taking timestamps at the point in time where \gls{hrtc} receives the first \gls{wfs} data and timestamps at the point in time where the last command data departs the \gls{hrtc}.

\begin{figure}[ht]
  \begin{center}
    \includegraphics[width=0.7\columnwidth,
                     clip,
                     trim= 0.4cm 0cm 0.5cm 1cm]{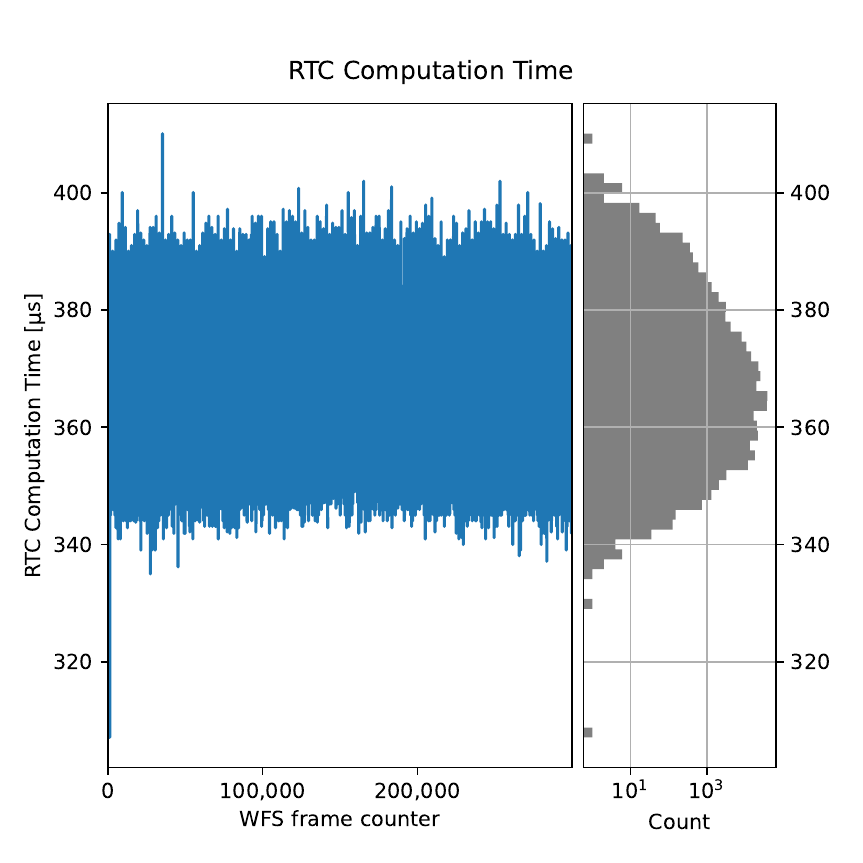}
  \end{center}
  \caption[RTC computation times in performance experiment]
  {\label{fig:hrtc_rtc_comp_time}
  RTC computation time series with its histogram in the performance experiment 
  for 300,000\,WFS images.
Median: 365\,\textmu{}s; 
mean: 365\,\textmu{}s; 
standard~deviation: 7\,\textmu{}s,
min: 307\,\textmu{}s; 
max: 410\,\textmu{}s; 
99.99\% percentile: 396\,\textmu{}s.
        }
\end{figure}

Figure~\ref{fig:hrtc_rtc_comp_time} shows the RTC computation time series. The result is that \gls{hrtc} was able to process each of the \gls{wfs} images in less than 420\,\textmu{}s.

During science observations, the \gls{srtc} will update parameters on the \gls{hrtc} on a regular basis. In order to assess the performance degradation during parameter updates, the \gls{srtc} updates the command matrix on the \gls{hrtc} every two seconds. 
Since the command matrix is the largest parameter with a size of about 249\,MB,  its update will have the biggest impact on the \gls{wfc} loop performance.


The parameter update consists of two steps: first, the \gls{srtc} sends the \gls{cm} to the \gls{hrtc} which then copies it to a passive parameter bank in the memory of the GPUs.
In addition to the passive parameter bank, the \gls{hrtc}  has an active parameter bank in the GPU memory which the \gls{wfc} control loop uses during operation. 
After the \gls{cm} has been uploaded to the GPUs,
the \gls{srtc} instructs the \gls{hrtc} to switch between the two parameter banks. As a result, the former passive parameter bank becomes the active one.

The \gls{cm} transfer from the \gls{srtc} into the \gls{hrtc} GPU memory takes about 0.8~seconds.
The time to copy the \gls{cm} from \gls{hrtc} main memory to GPU memory is roughly 0.3\,s. 
Switching between the active and passive parameter banks takes less than 0.1\,s from the initiation of the switch by the \gls{srtc} until the acknowledgement is received.

The \gls{metis} \gls{hrtc} transmits data with a total throughput of 306\,MB/s into the network. The majority of that throughput is caused by MUDPI telemetry data with about 286\,MB/s. In total, this \gls{hrtc} receives data with a throughput of approximately 168\,MB/s. Here, the \gls{wfs} image stream dominates that input throughput with roughly 148\,MB/s. 


\subsubsection{Numerical Correctness Check} 

\begin{figure}[ht]
  \begin{center}
    \includegraphics[width=0.9\columnwidth]{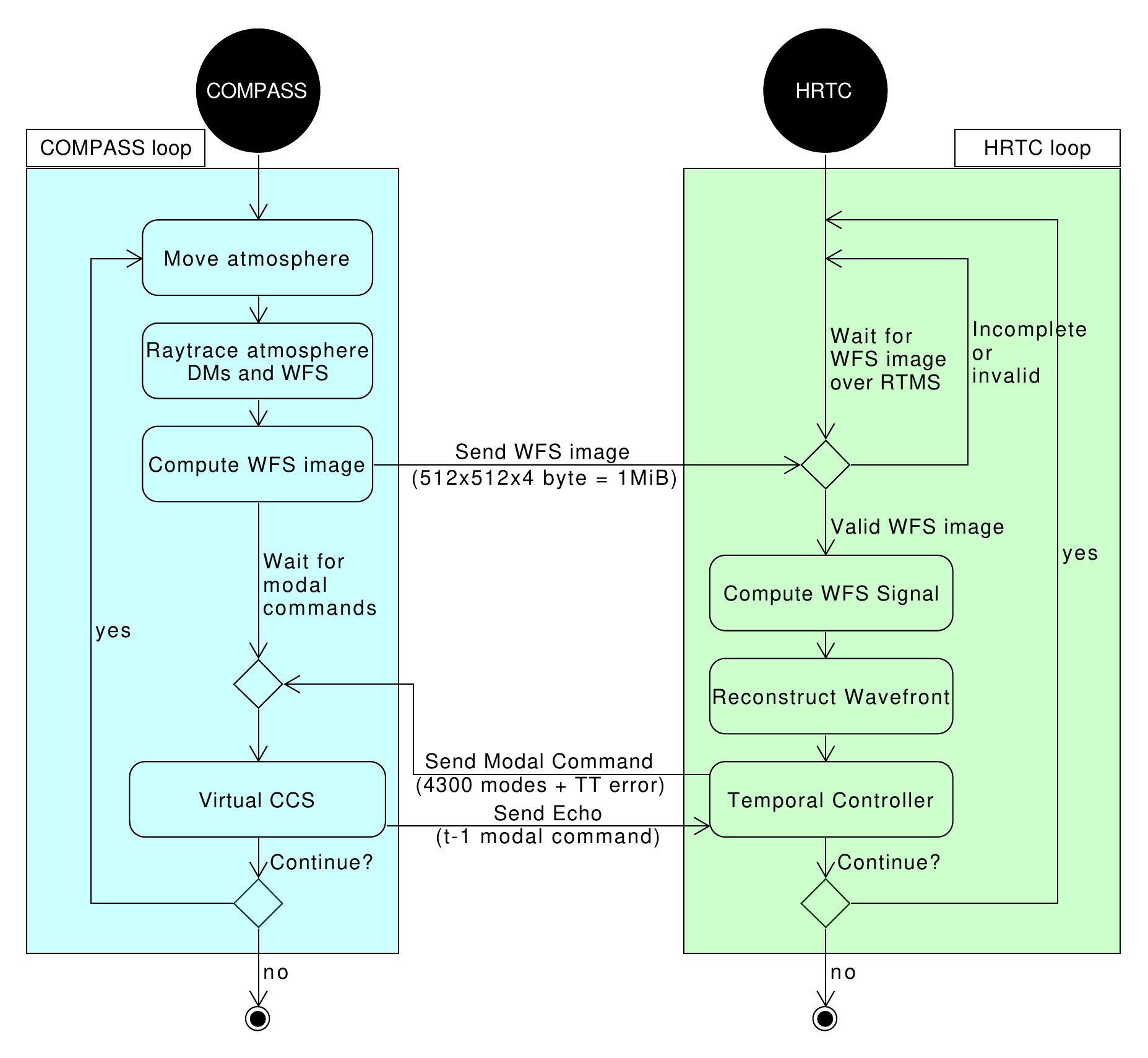}
  \end{center}
  \caption[COMPASS HRTC Integration diagram]{\label{fig:COMPASS_HRTC}
  Flow diagram illustrating the integration of COMPASS and \gls{hrtc} in the numerical correctness check.}
\end{figure}

An experiment was conducted to verify the numerical correctness of the \gls{hrtc} prototype. 
To this end, the \gls{hrtc} was integrated to run 'in the loop' of the COMPASS simulation described in Section~\ref{sec:simulation-setup}, essentially replacing the COMPASS in-built RTC functionality as show in figure~\ref{fig:COMPASS_HRTC}. Thus, COMPASS is responsible for simulating the atmosphere, calculating the forward model of the \gls{pwfs} and producing the simulated \gls{wfs} image. 
COMPASS was extended to send the  WFS images via \gls{rtms} to the \gls{hrtc},  in effect
it replaced the WFS camera simulator of the performance experiment.

The HRTC WFC pipeline computes the modal mirror commands and sends them to COMPASS. After COMPASS has received these commands,  it applies them to a virtual \gls{ccs} (see Sec.~\ref{sec:baseconfig}). 
Finally,  COMPASS returns an echo to the \gls{hrtc} containing the actually applied command.

The numerical correctness check was successful because the simulations that incorporated the \gls{hrtc} were completed with the expected performance regarding the quality of the wavefront reconstruction \citep[c.f.][]{feldt23}.  This test verifies that the \gls{hrtc} \gls{wfc} pipeline is numerically correct.

\subsection{SRTC}\label{sec:SRTC}

The \gls{srtc} oversees the operation of the \gls{hrtc} which includes the estimation of system parameters, control loop optimisation and other diagnostic and miscellaneous tasks. Table~\ref{fig:srtc-table} shows some of the most important functionality of the \gls{srtc}.

\begin{table}[htpb]
  \footnotesize
  \caption{Loop co-processing and optimisation tasks}
  \label{fig:srtc-table}
  \centering
  \begin{tabular}{p{\dimexpr 0.3\columnwidth - 2\tabcolsep}p{\dimexpr 0.70\columnwidth - 2\tabcolsep}}
  \toprule  
  \textbf{Co-processing Requirement} & \textbf{Functionality} \\
  \midrule
  System Parameter Estimation &  
    \vspace{-1em}
    \begin{itemize}[noitemsep, topsep=0pt]
      \item Pupil position monitoring (lateral motion and clocking) 
      \item Mis-reg identification 
      \item Interaction matrix calibration
      \item Valid sub-apertures map 
      \item Valid actuators map 
      \vspace{-\baselineskip}
    \end{itemize} 
    \\
    \midrule
  Control loop parameter optimisation & 
    \vspace{-1em}
    \begin{itemize}[noitemsep, topsep=0pt]
        \item  Control parameter optimisation 
        \item  Control matrix optimisation
      \vspace{-\baselineskip}
    \end{itemize}  \\
  \midrule 
  Offsets, Statistics and Diagnostics &
    \vspace{-1em}
    \begin{itemize}[noitemsep, topsep=0pt]
        \item \gls{psf}, Strehl ratio, \gls{fwhm}
        \item Contrast 
        \item Pupil fragmentation 
        \item PSDs (temporal, spatial)
        \item Telemetry statistics 
        \item Saturation statistics
      \vspace{-\baselineskip}
    \end{itemize} \\
  \bottomrule
 \end{tabular}
\end{table}

\begin{figure}[ht]
  \begin{center}
    \includegraphics[width=0.8\columnwidth]{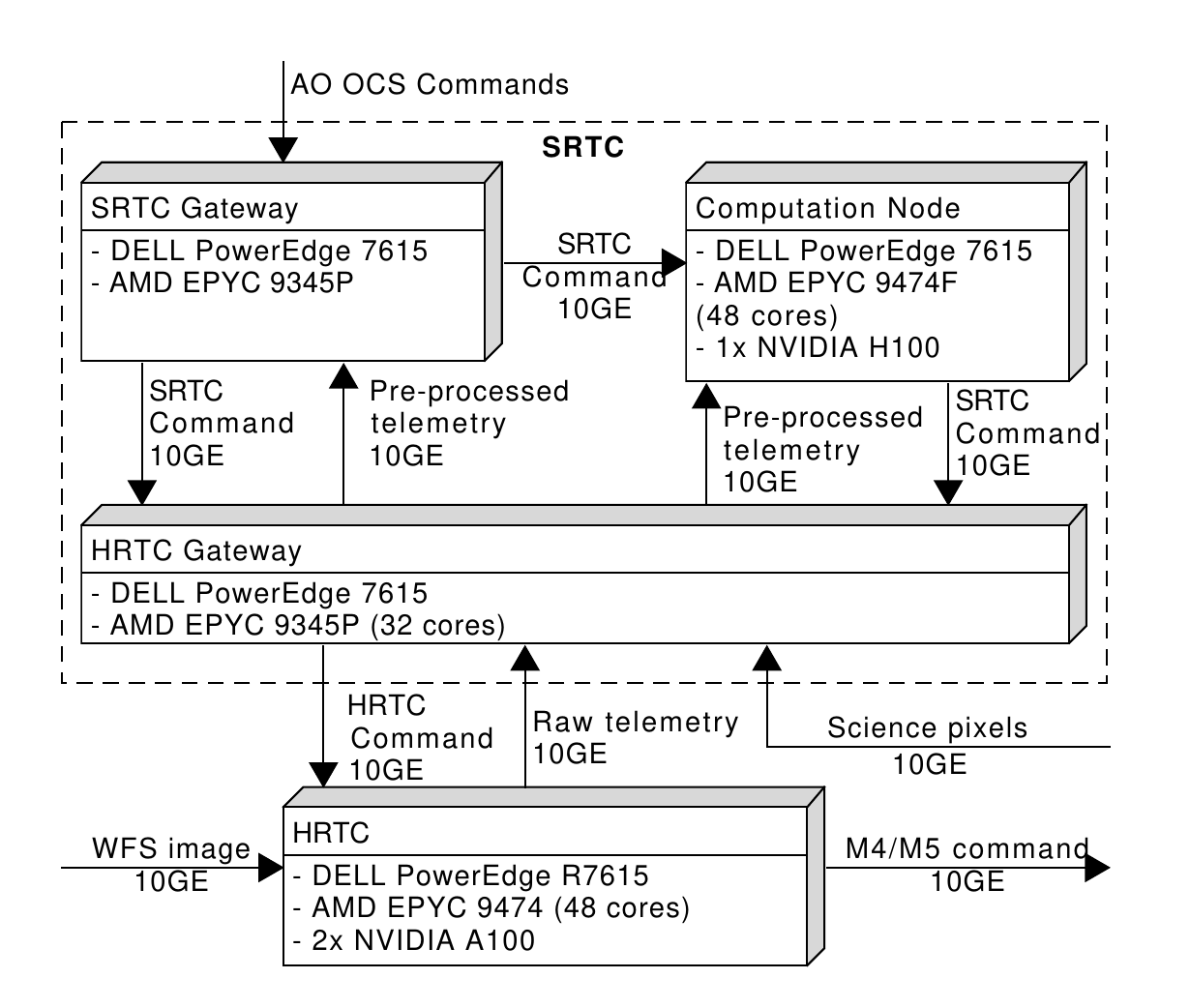}
  \end{center}
  \caption[RTC communication]{\label{fig:RTC_Communication}
  The \gls{rtc} computer hardware structure and its communication paths 
  inside of METIS \gls{aocs}.
  }
\end{figure}

The \gls{srtc} will be a distributed computer system.  Its foreseen computer nodes, the communication paths and the \gls{srtc} environment are depicted in figure~\ref{fig:RTC_Communication}. We plan to use the ESO IT standard computers to facilitate maintenance at the \gls{elt}.

While the \gls{hrtc} is being fully developed in-house at the \gls{mpia}, the \gls{srtc} will be developed using the \gls{rtc} Toolkit. The \gls{rtc} Toolkit is a suite of software tools and libraries, developed and maintained by \gls{eso}, that provides a scalable framework onto which instrument-specific application code and configuration are added. In essence, it provides a skeleton for the implementation of the \gls{srtc} components.

\subsubsection{Components}

\begin{figure}[ht]
  \begin{center}
    \includegraphics[width=0.7\columnwidth]{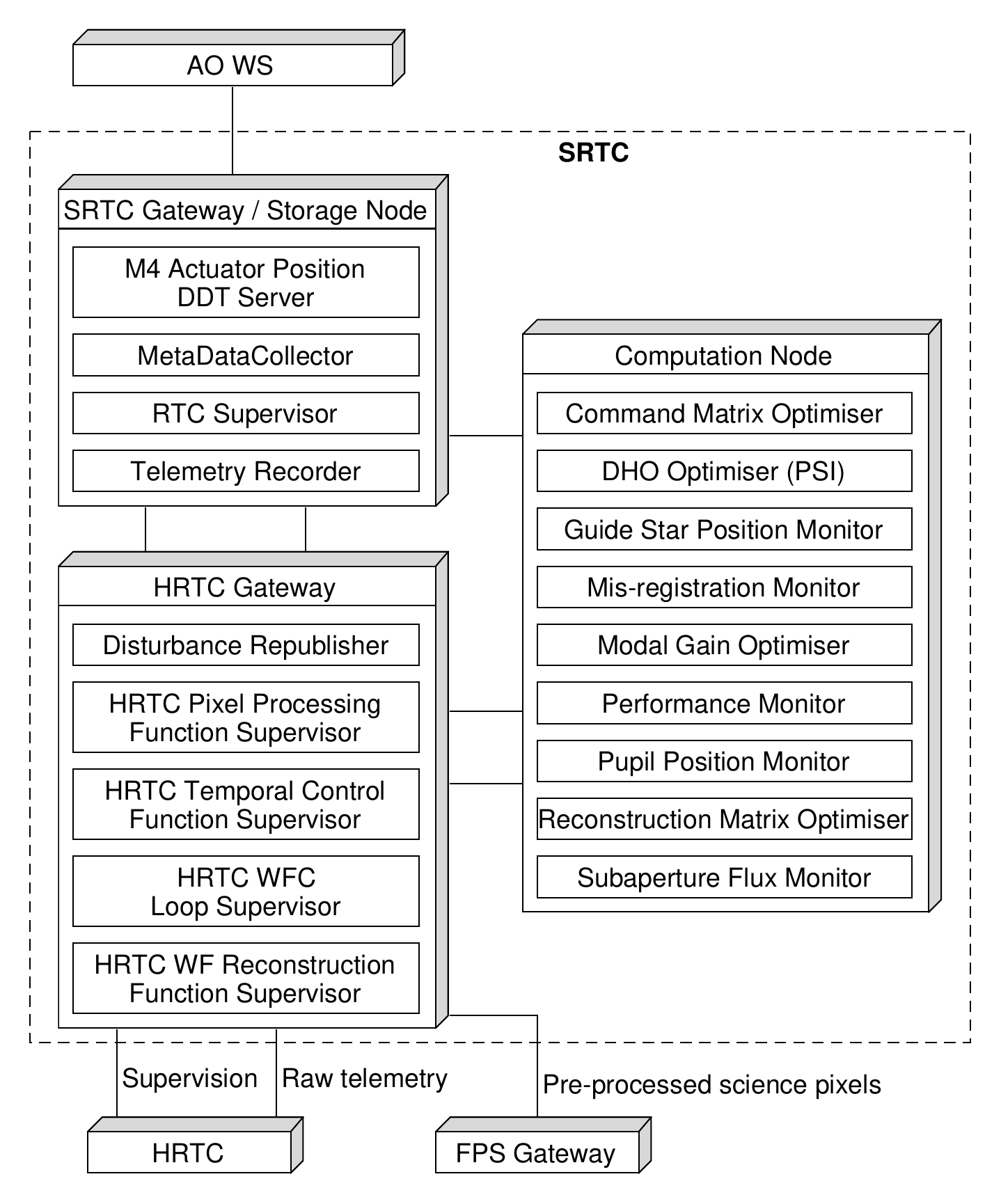}
  \end{center}
  \caption[RTC communication]{\label{fig:SRTC_components}
  SCAO \gls{srtc} software deployment. This diagram shows the expected \gls{srtc} components deployed on each cluster node.}
\end{figure}

The \gls{srtc} is a distributed system, running on three nodes with specific use cases. Figure \ref{fig:SRTC_components} shows the deployment diagram for the \gls{srtc} nodes and the expected interconnect between each of the three nodes. All \gls{srtc} nodes will run the \gls{rtc} Toolkit (RTC Tk), while implementing each of the components shown. 

The \gls{srtc} Gateway provides access to the RTC functionality towards higher-level control systems such as METIS ICS. The Storage Node is equipped with large and fast persistent storage in order to record WFC loop telemetry data for offline analysis, and is currently envisioned to be unified with the \gls{srtc} Gateway node. Although not yet fully defined, the Storage Node is expected to handle more than 500\,MB/s of telemetry and meta-data. 

The \gls{hrtc} gateway is the gateway between the \gls{hrtc} and the \gls{srtc} nodes. It has two tasks. The first task of this node is to act as a gateway for telemetry data and disturbance data. In this case, it can be called \gls{hrtc} Telemetry Gateway and \gls{hrtc} Disturbance Gateway. The other task consists of supervision of the \gls{hrtc} which results in the name \gls{hrtc} Supervision Gateway. Both of the tasks requires a fast connection to the telemetry network.

The most demanding node will be the Computation Node, and will use GPU accelerators for raw compute power. The Command Matrix Optimiser (CMO) is expected to be the most demanding process running on this node, with an expected minimum interval of 2 seconds between subsequent executions when observing near zenith.

\FloatBarrier


\section{Simulating SCAO \label{sec:scao-sim}}

Simulations of \gls{metis}' \gls{scao} system served three purposes in particular:

\begin{enumerate}
    \item To facilitate the choice of fundamental parameters of our \gls{scao} system, such as the ones listed in Sec.~\ref{sec:design_decisions}
    \item To evaluate the robustness of the system with respect to a number of adverse effects, such as wind-induced vibrations, \gls{ncpa}s, fabrication errors of the pyramid prism and the like.  Permanently present ones enter the error budget, transient or variable ones simply undergo a robustness analysis.
    \item To predict the performance of the system and verify requirement compliance.
\end{enumerate}

One additional step we took is the so-called superhero simulation, which assesses the impact of combined errors in a single simulation, instead of quantifying the impact of each individual one.

\subsection{Simulation Setup \label{sec:simulation-setup}}

\begin{table}[ht]
\centering

\caption[Standard Parameters]{Standard Parameters}

\label{tab:stdpars}

\begin{tabular}{lccl} 
\hline
 Parameter  & Value & comment \\ \hline
 Seeing  & 0\farcs66  & (@0.5\,$\mu$m, median) \\
 Zenith angle & 30$^\circ$ &  \\
 \hline
 Simulation length & 60\,s &  \\
 Loop frequency & 1\,kHz & \\
 \hline
 Telescope aperture & ELT & 54\,cm spiders \\
 \hline

 Wind on telescope & 8\,m/s & from 20$^\circ$ off pointing \\
 Optical throughput & 0.2 & including QE\\
 Detector dead time & 272\,$\mu$s\\

 \hline
 Science wavelength & 3.7\,$\mu$m & \\
 Guide star magnitude & 4.537 & m$_K$ \\
 \hline
 RTC delay & 1.1 & frames (at 1\,kHz) \\
 tt ctrl. P gain & 0.3 & tip-tilt\\
 tt ctrl. I gain & 0.29 & tip-tilt\\
 hm ctr. P gain & 0 & high-order \\
 hm ctr. I gain & 0.45 & high-order \\

\hline
\end{tabular}

\end{table}





\subsubsection{COMPASS Modules \label{subsubsec:compass_modules}}
Simulations were carried out using the COMputing Platform for Adaptive optics SystemS (\gls{compass}) environment \citep{github_compass} in version 5.0, a tool actively under development at LESIA \citep{2016SPIE.9909E..71G}. \gls{compass} has been extended by additional modules called {\tt p\_metis}, {\tt p\_calibration\_DM}, and {\tt p\_ccs}.

The first two modules essentially implement our reconstruction scheme using the virtual Fried geometry DM and the regularised Minimum Mean Square Error (\gls{mmse}) inversion and projection steps for the full set of available modes on M4 \citep{obereder23}. The {\tt p\_metis} module also includes features for operating on extended sources, applying NCPAs, dealing with nonideal pyramid prism shapes, dynamic M1 segment aberrations, and more.

The {\tt p\_ccs} module takes the generated modal command vector, facilitates the collaborative control scheme between \gls{m4} and the ELT's tip-tilt field stabilisation mirror \gls{m5}, while performing saturation avoidance checks as described in \cite{ESO-311982-V3}. For more details see Sec.~\ref{sec:ccs}. 

\subsubsection{Baseline Configuration at FDR \label{sec:baseconfig}}

Throughout the different phases of the project, we used so-called baseline configurations, and deviations from these baselines were made only for specific analyses in small subsets of the parameter set. The main baseline configuration for phase D specifies a bright star in median atmospheric conditions at 30 degrees zenith distance seen through an ELT pupil with 54 cm spiders. Our standard simulation runs for 60 seconds at 1 kHz. A summary of the setup can be seen in Tab.~\ref{tab:stdpars}.

The reconstructor utilises a \gls{vdm} tied to the \gls{pwfs}'s pixel grid in a Fried geometry. The \gls{vdm} uses bilinear ansatzfunctions as influence functions ("pyramidlets"), something a physical \gls{dm} never could. The virtual calibration is done with an actuator push amplitude (aka {\tt push4imat}) of 50\,nm to stay well within the sensor's linear regime. The reconstructed wavefront is projected onto a set of force-aware Karhunen-Loeve modes \citep{correia23}. We always project onto and control the full set of modes, with higher-order modes receiving sequentially smaller amplitudes via the regularisation mechanism \citep{obereder23}.

Table~\ref{tab:stdpars} describes the setup of the PI controller for high-order and tip-tilt control. The {\tt P\_gain} and {\tt I\_gain} factors $g_I$ and $g_P$ translate into control commands according to:

\begin{equation}
\mathbf{m}_k = \gamma \mathbf{m}_{k-1} + (g_I + g_P)\mathbf{e}_k - g_P \mathbf{e}_{k-1},
\end{equation}
where $\mathbf{m_k}$ denotes a modal command at loop step $k$, $k$ is the loop index, and $\mathbf{e_k}$ is the error signal at loop step $k$. $\gamma$ is the leakage factor with $0 < \gamma \leq 1$ in theory, but $\gamma$ is very close to 1 in practice.



\subsubsection{The Central Control System\label{sec:ccs}}
The \gls{ccs} module is set up according to \cite{ESO-311982-V3}. The module enforces limits on the maximum actuator gap (30\,$\mu$m), maximum speed (2.5\,m/s), and maximum force exerted by each actuator (1.2\,N). 

\gls{ccs} enforces these limits in a dedicated scheme:

\begin{enumerate}
    \item Calculate a \gls{dm} shape from the first 50 modes, and check for gap (30\,$\mu$m) and speed (2.5\,m/s).  If either limit is violated, scale the modal command vector in order to scale the requested wavefront accordingly.
    \item Check the maximum applied actuator force. If more than 1.2\,N are requested from any actuator, cut the highest 40 modes from the command vector, and recalculate. Repeat until the maximum requested force is below 1.2\,N.
    \item Calculate the gap and speed on the resulting requested \gls{dm} shape. If either criterion is violated, drop the frame and re-apply the last command.
    \item Send an echo of the actual applied \gls{dm} shape to the controller.
\end{enumerate}
The \gls{ccs} also is in charge of offloading the tip and tilt components to \gls{m5} facilitating the collaborative control scheme of the two deformable mirrors \gls{m4} and \gls{m5}.

\subsection{Key Performance Indicators}

The standard metric to watch when simulating adaptive optics is, of course, the Strehl ratio. However, in \gls{metis}, the requirements placed on Strehl are quite relaxed and can be easily exceeded even under the worst conditions. The true challenging requirements are on the final contrast achieved, and on residual image motion and petal piston. Therefore, we use the Adaptive Optics Simulation Analysis Tool(kit) (\gls{aosat}\footnote{\gls{aosat} can be obtained from github (github.com/mfeldt/AOSAT), the documentation is on readthedocs (aosat.readthedoccs.io).}) package \cite{feldt20} for the analysis of \gls{scao} simulation results. \gls{aosat} provides a suite of "analyzers" for our \glspl{kpp}.


The setup used for most analyses includes analysers for the \gls{psf} (Strehl ratio, residual TT jitter), the Temporal Variance Contrast (\gls{tvc}), pupil fragmentation (petal piston), and Zernike decomposition of the residual wavefronts. 

\subsubsection{Strehl and Residual Tip Tilt}

The Strehl ratio is calculated in two ways, once from the residual wavefront deviations, and from the peak intensity of the \gls{psf} compared to a perfect \gls{psf} calculated from a flat wavefront.  
For further evaluations, we generally use the Strehl ratio computed from the peak intensity.  Residual image jitter can also be computed either from the wavefront or from the image directly, for analysis we generally use the actual \gls{psf} motion.

\subsubsection{Petal Piston}

Wavefront piston and tip-tilt is evaluated individually on each of the six petals of the \gls{elt} for each of our simulations.    To track the performance of our system in terms of petal piston when varying a given parameter, we generally plot the maximum of the six rms values against said parameter.

\subsubsection{Temporal Variance Contrast}

\begin{figure}[htb]
\centering
\includegraphics[width=.65\textwidth]{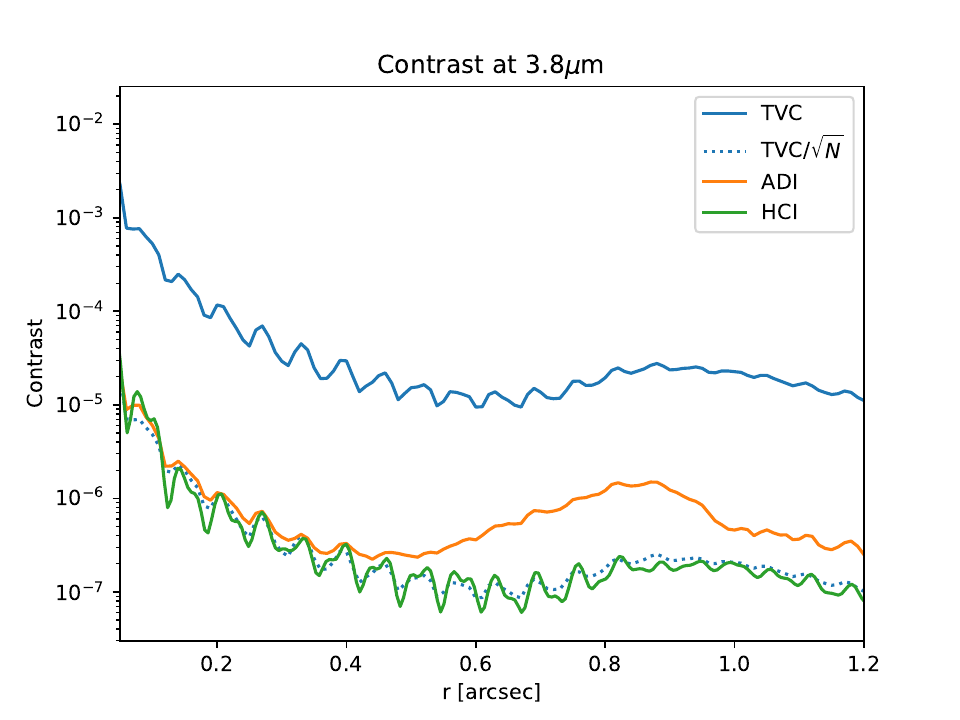}
\caption{Comparison of the achieved contrast as derived by our TVC metric (blue curves), and a full fledged \gls{hci} simulation (green).}
\label{fig:contrast_tvc_vs_hci}
\end{figure}

The \gls{tvc} is an estimate of the achievable $5\sigma$ contrast using a "perfect" coronagraph versus separation.
The estimation is, as the name implies, derived from the temporal variance of the \gls{psf}. Since at this stage our simulations do not include aberrations that change on timescales between seconds and hours,
all residual variation is from the atmosphere, or the \gls{ao} control only.  This state represents a perfect subtraction of whatever additional aberrations might cause slowly changing \gls{psf} structures
by an idealised kind of differential imaging method.  The resulting \gls{psf} variations thus represent the ultimate limitation to the achievable contrast, and can be scaled to any desired observation length. 

This temporal variation is calculated on each detector pixel from the full set of \gls{psf}s derived in previous steps, and plotted versus the angular separation of the pixel concerned from the image centre.
Note that the calculation is done per pixel, where in practice one would measure per airy disk. 

At the time of \gls{pdr} we conducted a dedicated study to compare the contrast curve derived by the \gls{tvc} method and a full-fledged simulation of \gls{hci} including dedicated coronagraphs and full post-processing. The outcome is shown in Fig.~\ref{fig:contrast_tvc_vs_hci}.  The green curve denotes the result derived by the full 1 hour \gls{hci} simulation, the orange curve is from a simplified ADI model, and the blue curve was derived by our \gls{tvc} method.  The blue dotted curve is scaled by the square root of the ratio of the number of uncorrelated phase screens in the simple \gls{tvc} simulation, and the number of phase screens in the full \gls{hci} simulation.   Although the excellent match between \gls{hci}-derived and \gls{tvc}-derived contrast is easy to understand because in the absence of mid- and long-term variations of aberrations the remaining temporal variance of the \gls{psf} \textit{is} what ultimately limits the achievable contrast, we generally do not make the effort of scaling the \gls{tvc} because we are only interested in the overall evolution when changing \gls{scao}-relevant system parameters.

\subsection{Baseline Performance}


 

A full suite of simulations was run to determine the baseline performance, as can be expected under various conditions in the absence of additional error sources. In order to adapt to changing guide star magnitudes and atmospheric conditions, we allowed the loop frequency and the regularisation parameters to vary between selected discrete values.  Note that both regularisation parameters, for reconstruction and projection, were always kept identical.



\subsubsection{Guide Star Magnitude}

\begin{figure*}[htb]
\centering
\includegraphics[width=0.49\columnwidth]{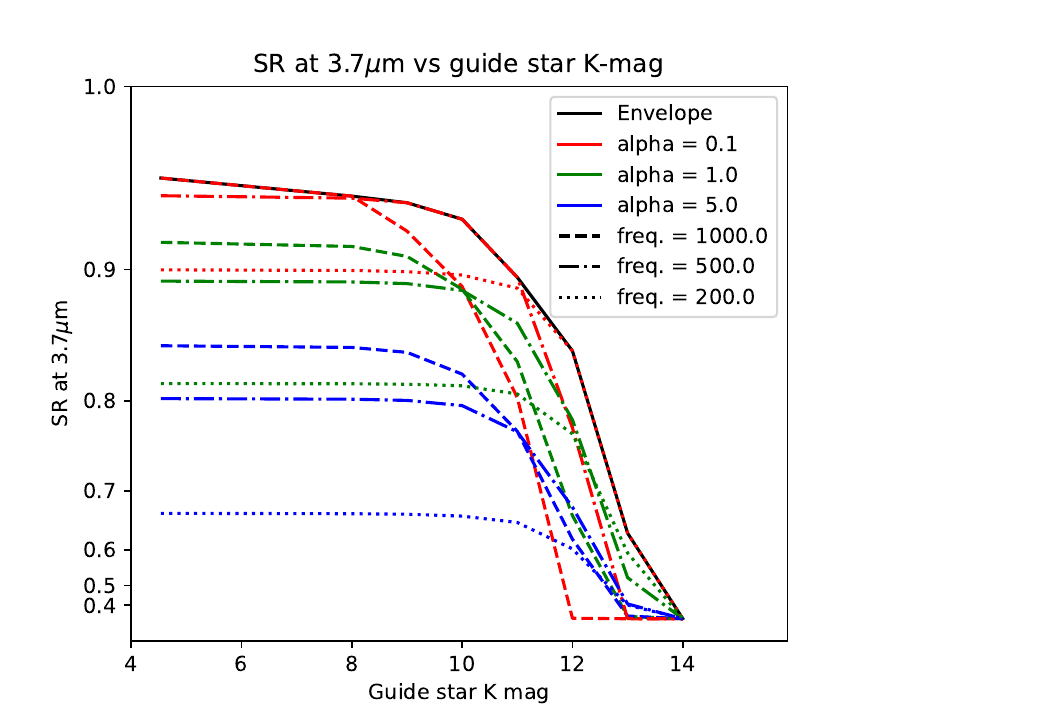}
\includegraphics[width=0.49\columnwidth]{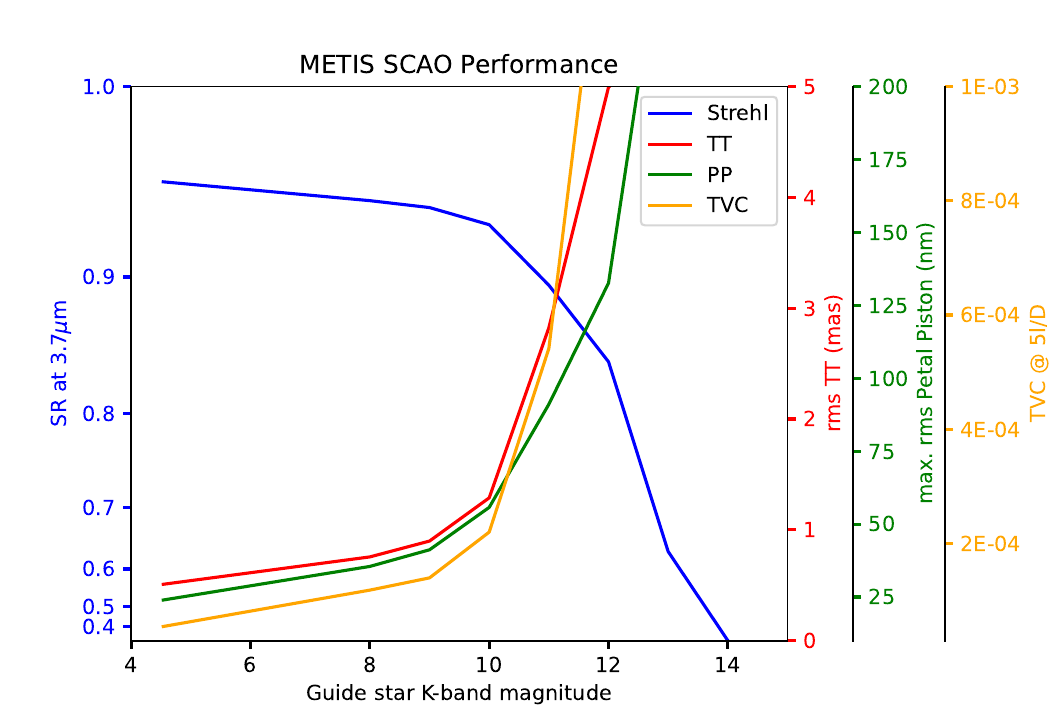}
\caption{\gls{scao} performance as a function guide star $K$ magnitude. The simulations are done for median seeing conditions a zenith distance of $30^\circ$. The loop frequency and regularisation strengths were varied according to the values mentioned in the legend.  regularisation values were always identical for reconstruction and projection. The left plot shows the performance in terms of Strehl number individually for each setting of regularisation factor and loop frequency. The right plot shows all our Key Performance Indicators (\gls{kpp}s), each one for an optimized combination of regularisation factors $\alpha_\mathrm{rec} = \alpha_\mathrm{proj}$ (See Eq.~\ref{eq1}) and frequency for each magnitude. }
\label{fig:sim-dFull-perf-mag}
\end{figure*}

Fig.~\ref{fig:sim-dFull-perf-mag} shows that we can achieve a limiting magnitude for our \gls{scao} system of about 13\,mag in $K$-band by decreasing the loop frequency to 200\,Hz.  When seeing and zenith distance are kept constant, changing the regularisation adds no additional advantage in faint conditions.  Note that the highest loop frequency of 1\,kHz delivers the best performance only down to a guide star magnitude of $m_K=$8\,mag.  Beyond that, 500\,Hz is the best frequency in our set, and beyond 11\,mag, the frequency is best reduced to 200\,Hz.

\subsubsection{Seeing Condition}

\begin{figure*}[htb]
\centering
\includegraphics[width=0.49\columnwidth]{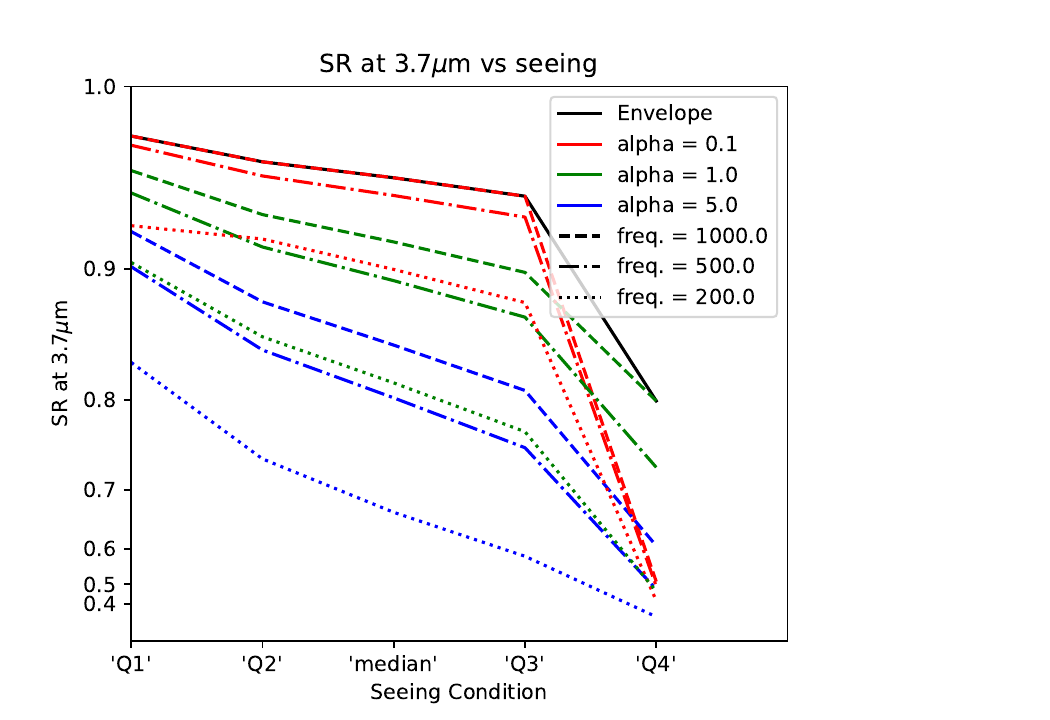}
\includegraphics[width=0.49\columnwidth]{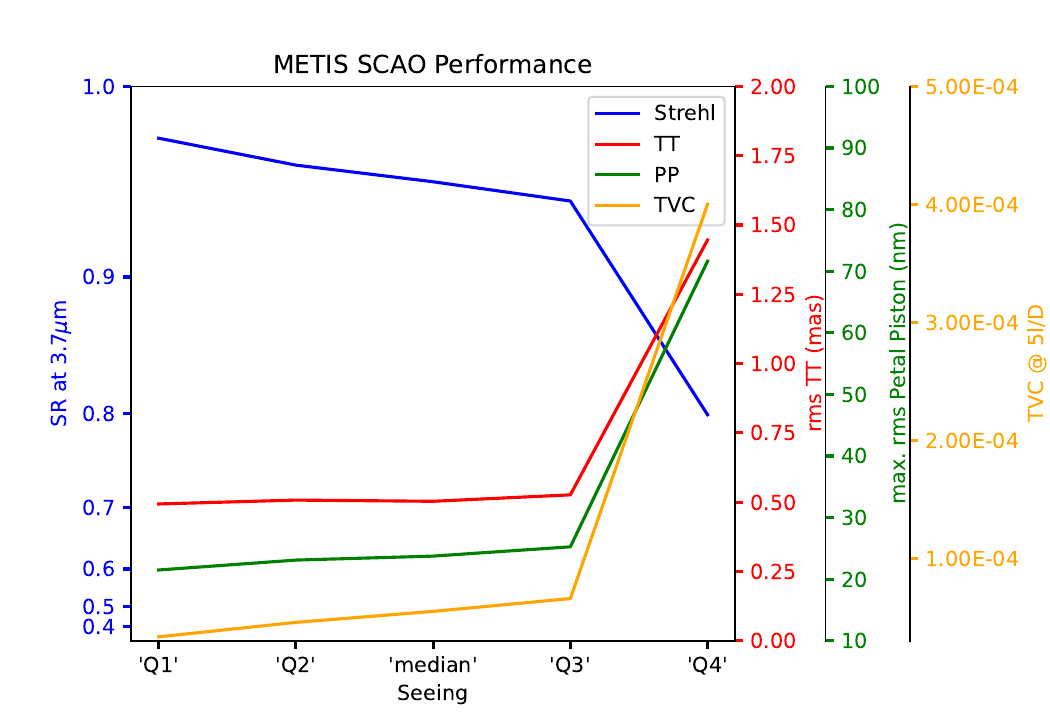}
\caption{\gls{scao} performance as a function seeing. The simulations are done for a bright star (m$_K$=4.5\,mag) at a zenith distance of $30^\circ$. The loop frequency and regularisation strengths were varied according to the values mentioned in the legend.  regularisation values were always identical for reconstruction and projection. The left plot shows the performance in terms of Strehl number individually for each setting of regularisation factor and loop frequency. The right plot shows all our \gls{kpp}s, each one for an optimized combination of regularisation factor $\alpha_\mathrm{rec}=\alpha_\mathrm{proj}$ (see Eq.~\ref{eq1}) and frequency for each magnitude.}
\label{fig:sim-dFull-perf-seeing}
\end{figure*}


The performance of \gls{scao} is roughly stable throughout 60\% of the time, that is, up to Q3. For Q4 seeing conditions, i,e, conditions that occur during the second worst 20\% of time on Cerro Armazones, it is favourable to increase the regularisation factor. From our available set, a regularisation factor of 1.0 for both reconstruction and projection delivers the best performance here. Note that since simulations are performed for the standard case of a $m_K=10$\,mag star, the best performance is always seen for a loop frequency of 500\,Hz.

\subsubsection{Zenith Angle}


Variations of the \gls{scao} performance with zenith angle expectantly behave like a mild version of the variation with seeing.  For zenith distances higher then $z=45^\circ$ in median seeing conditions the regularisation factors should be increased, as they should for seeing conditions worse than Q3 at $z=30^\circ$. 

\subsubsection{Overall Performance}
 All in all we note that \gls{metis}' \gls{scao} delivers an excellent and exceptionally robust performance. We expect a peak performance in excellent seeing conditions on bright stars near zenith (not shown in any Figure) of $S_{3.7}$=97.8\% Strehl at 3.7\,$\mu$m, corresponding to a residual wavefront error rms of 88\,nm. Performance under "nominal conditions", that is, on a star of m$_K=4.5$\,mag, 30$^\circ$ from zenith in median seeing and including all independent error sources corresponds to a Strehl ratio at 3.7\,$\mu$m of $S_{3.7}=$95.4\%, or a residual wavefront error of 128\,nm.  
 
 We can operate the system almost as a "pushbutton" \gls{ao}, the only parameters adapting to changing conditions being the loop frequency when stars get fainter than about m$_K=8$\,mag, or the regularisation when the seeing conditions become very bad. The limiting magnitude up to which a noticeable, decent performance can be expected is determined to be near m$_K=13$\,mag where we expect a performance of $S_{3.7}=63.6$\% at $z=30^\circ$ in median seeing conditions, corresponding to a residual wavefront rms of 396\,nm.

 Be aware that the numbers quoted describe the performance of the \gls{scao} system alone under the simulated baseline conditions, not of the entire \gls{metis} instrument. The overall performance including the handling of additional, "real-world" error sources in day-to-day operations is described in Sec.~\ref{sec:superhero}.

\subsection{Robustness Analysis}
\subsubsection{Example: NCPA and Water vapor Seeing\label{sec:sim-ncpa}}

\begin{figure}[htb]
\centering
\includegraphics[width=.55\textwidth]{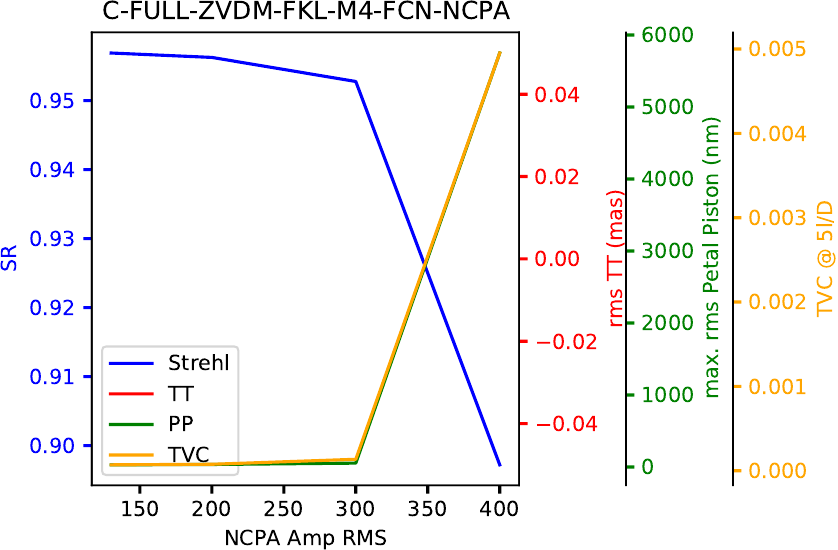}
\caption{\gls{scao} performance as a function of applied \gls{ncpa}s. A number of optimisations were applied, see text.}
\label{fig:C-FULL-ZVDM-FKL-M4-FCN-NCPA}
\end{figure}

In METIS, optical NCPAs were originally expected to be below 130\,nm, but current NCPA maps from the optical design only show values up to 92\,nm. Being conservative, we still use the 130\,nm threshold as the expected value for NCPAs.

In this experiment, however, we included residual water vapor (WV) seeing, effectively introducing NCPAs between WFS and science path which follow a Kolmogorov-like power spectrum. In order to achieve this, 100 Zernike terms were fitted to an open-loop atmospheric phase screen, and the result was used as optical NCPAs. The rms from water vapor seeing is expected to be of the order of 300\,nm, reaching 400\,nm under bad conditions. Thus, we simulated four NCPA rms values of 130 (the upper limit for expected NCPA purely from optics), 200, 300, and 400\,nm.

In addition, we included wind-induced tip-tilt according to ESO-supplied power spectra for a wind speed of 8\,m/s blowing from 20$^\circ$ off the pointing direction.

The system is not easy to stabilize under the presence of both NCPA and wind-induced tip-tilt. In order to find a stable performance, a number of setup parameters were allowed to vary, most notably the modulation amplitude, the regularisation factors (which were both, for reconstruction and projection, kept identical), and the loop gain (in this particular experiment, the controller was a pure integrator).

\begin{table}[htb]
\centering
\caption[WV NCPA Optimized Parameters]{Reconstructor and Control Parameters optimized for each amount of WV-NCPAs. The numbers quoted produce the optimal Strehl ratio as measured on the PSF, the numbers in brackets produce the optimum TVC. The regularisation parameters for reconstruction and fitting $\alpha_\mathrm{rec} = \alpha_\mathrm{proj}$ were identically applied for reconstruction and projection of the wavefront.}
\label{tab:C-FULL-ZVDM-FKL-M4-FCN-NCPA}
\begin{tabular}{|c|ccc|} \hline
WV NCPA & Modulation & Reg. & control gain \\
rms [nm] & amplitude&  &  \\ 
    &   [$\lambda$/D] & $\alpha_\mathrm{rec}=\alpha_\mathrm{proj}$ & g\\ \hline
130 & 4 (4) & 0.02 (0.08) & 0.4 (0.5) \\
200 & 4 (4) & 0.02 (0.08) & 0.6 (0.5) \\
300 & 6 (6) & 0.02 (0.08) & 0.3 (0.6) \\
400 & 7 (7) & 0.32 (0.32) & 0.4 (0.4) \\
\hline
\end{tabular}
\end{table}

Fig.~\ref{fig:C-FULL-ZVDM-FKL-M4-FCN-NCPA} shows that we can indeed tolerate NCPAs of up to 300\,nm without significant loss of performance. At 400\,nm we can still operate the system in a stable manner, achieving a Strehl of 90\% at 3.7\,$\mu$m. Despite the fact that the water vapor seeing amplitude is growing with wavelength, we may still expect very good performance at 10\,$\mu$m. The best parameters to operate at each NCPA amount found are summarized in Tab.~\ref{tab:C-FULL-ZVDM-FKL-M4-FCN-NCPA}.

\subsubsection{Example: Low-Wind Effect}

\begin{figure}[htb]
\centering
\includegraphics[width=\columnwidth]{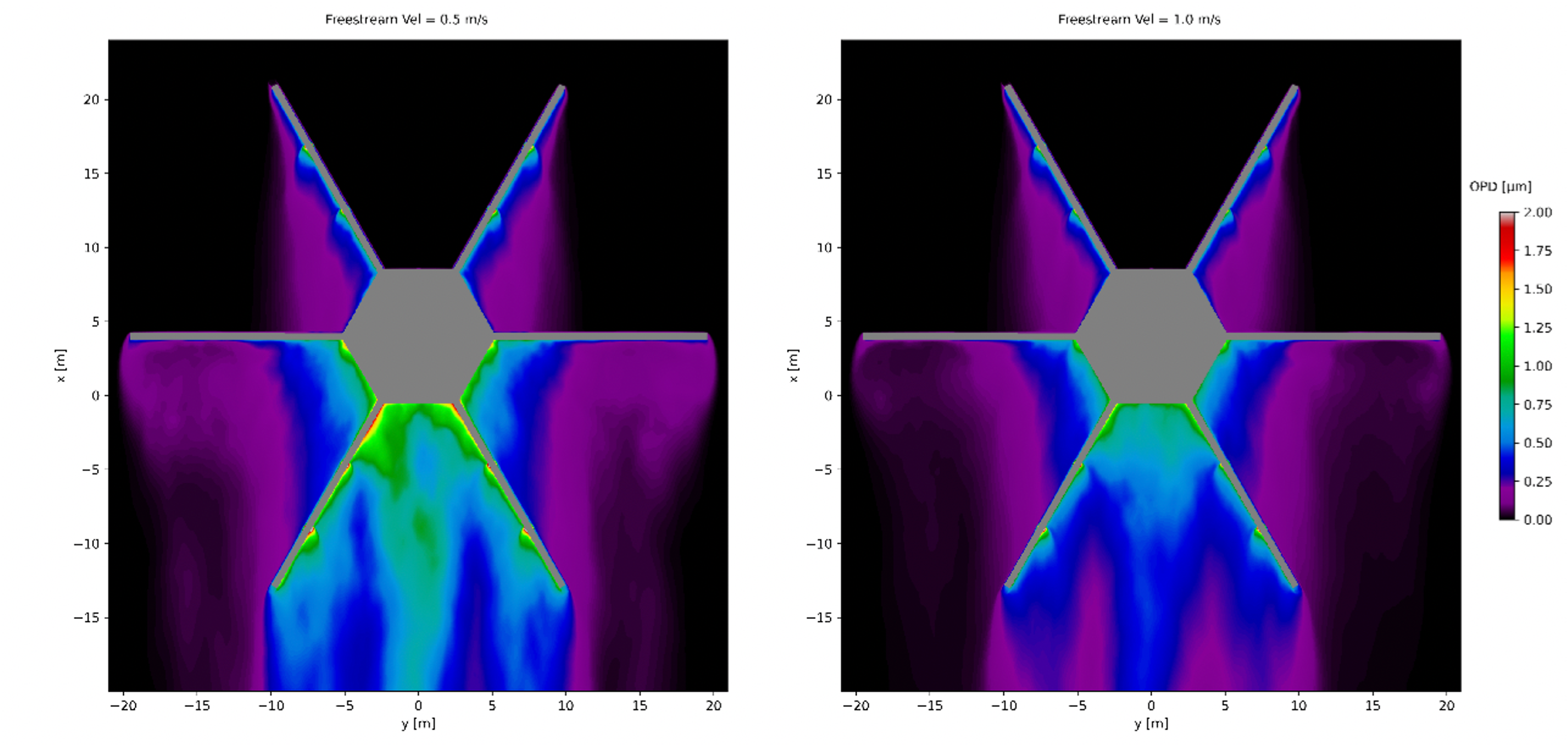}
\caption{Average Optical Path Difference (\gls{opd}) maps for 0.5\,m/s (left) and 1\,m/s (right) simulations. From \cite{martins22}}
\label{fig:lweMapsFromMartins22}
\end{figure}

During the night, the temperature of the girders of the secondary support structure can fall several Kelvin below that of the surrounding atmosphere due to radiative cooling \citep{couder49}.
Thus, in conditions of very low wind speeds, substantial temperature differences can occur between the air on one side of a girder and that on the other side. As a result, optical path differences of several microns between the two sides can be induced, difficult to sense and correct for any adaptive optics system. Frequently AO systems in such circumstances produce a petal pattern, where one fragment of the pupil delimited by support structures has a constant phase offset with respect to its neighbor, usually a multiple of the sensing wavelength.

\begin{figure}
\centering
\includegraphics[width=0.45\textwidth]{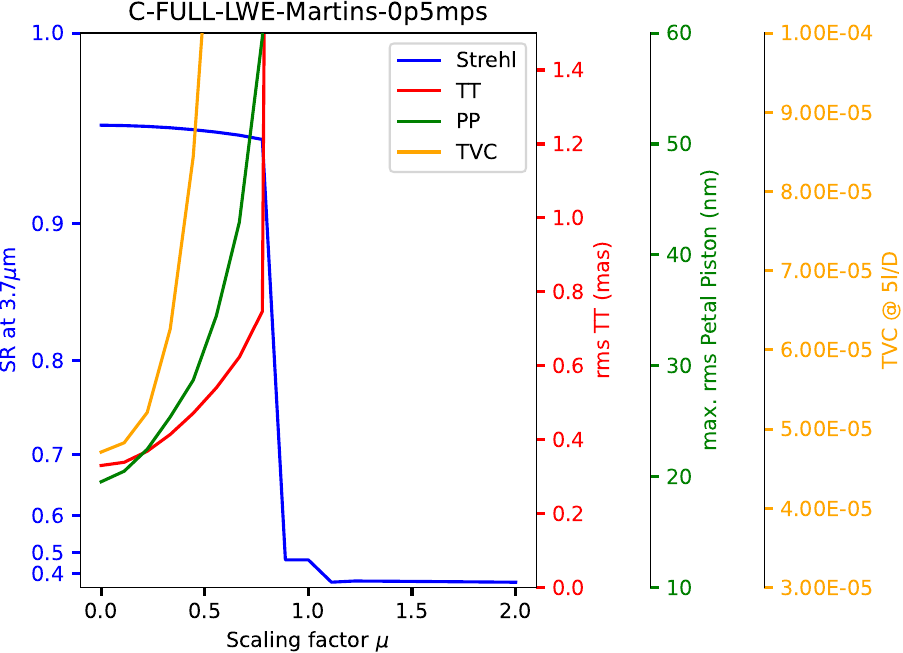}
\includegraphics[width=0.45\textwidth]{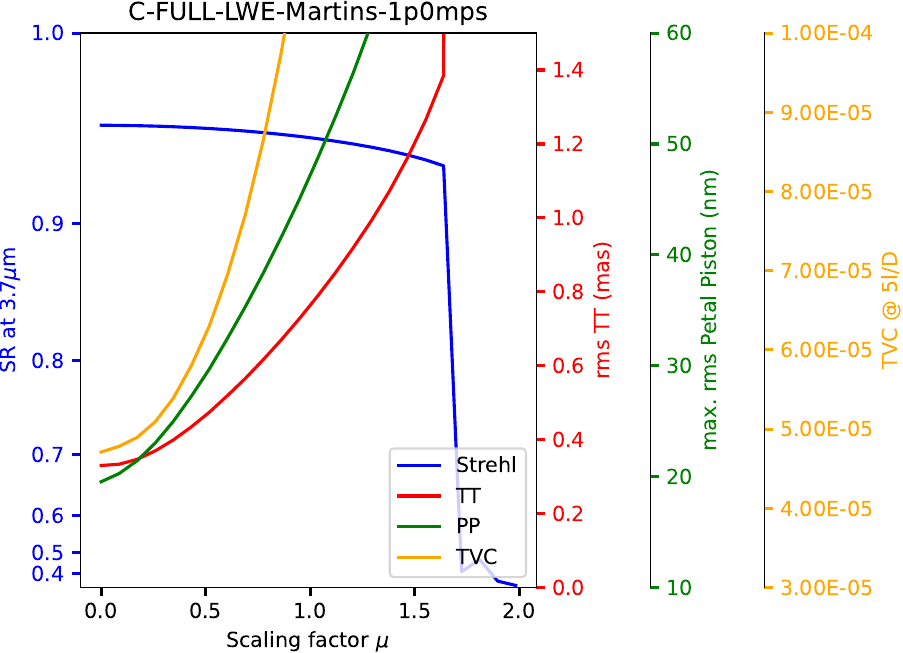}
\caption{Performance obtained in the presence of scaled low-wind-induced OPD maps for wind speeds of 0.5\,m/s (top) and 1.0\,m/s (bottom). The scaling factor $\mu$ is 1.0 where the nominal simulation results of \cite{martins22} are fully applied, including the scaling factor of 2.7 for the 0.5\,m/s and 1.74 for the 1.0\,m/s results.}
\label{fig:C-FULL-ZVDM-FKL-M4-FCN-LWE}
\end{figure}

The simulation the impact, we used  time-varying \gls{opd} maps obtained from computational fluid dynamic models for low-wind conditions \citep{martins22} in a standard simulation setup. Such maps were kindly provided by Ronald Holzloehner and are shown in Fig.~\ref{fig:lweMapsFromMartins22} for two low-wind conditions. Deviating slightly from the standard setup, a higher regularisation factor of 5.0 was applied in the reconstructor during an initial loop closure phase of 1200 frames. From experience, this helps to close the loop when the conditions are such that petalling is about to become problematic. Measurements of the standard parameter set as shown in Fig.~\ref{fig:C-FULL-ZVDM-FKL-M4-FCN-LWE} were made on the 6000 frames following this closure period.

Keep in mind that "standard setup" implies a bright star and median seeing conditions; the latter might actually not necessarily apply during low-wind conditions which are usually poised to occur during the most excellent seeing conditions.


Fig.~\ref{fig:C-FULL-ZVDM-FKL-M4-FCN-LWE} summarises the performance metrics for the impact of the LWE in our standard configuration. It is obvious that without further mitigation, we cannot run the system under conditions as simulated by \cite{martins22} for 0.5\,m/s in median seeing when these conditions prevail already at loop start-up. The system immediately locks into a 2$\pi$ petalling state and remains so. For 1\,m/s, the situation is better, and we could cope with nominal results ($\mu$=1) in the sense of being able to close the loop and run in a stable fashion, at least throughout our simulated 6s period. However, at $\mu$=1 there is already an impact on performance metrics, most notably a contrast loss of a factor of $\sim$2.5.

It should be mentioned that the OPD maps provided by \cite{martins22} are quite uncertain by the nature in which they were derived. For more information, see \cite{martins22}, where we can simply state that a couple of microns more or less in optical path difference are easily possible.

\subsubsection{Impact of Combined Errors (Super-Hero) \label{sec:superhero}}
\begin{table}[tbp]
\centering

\caption[Super Hero Errors]{Errors introduced in the super-hero simulation.}

\label{tab:super-hero}
\begin{tabular}{lc} 
\hline
 Error Term  & Amplitude   \\ \hline
 \textbf{Fixed Errors} & \\ \hline 
 M4 misrotation  & $0.2^\circ$\\
 M4 misalignment & 0.1\,m in x and y\\
 Missing segments & 21 in 3 flowers \\
 \textbf{Scaled Errors} & \\ \hline 
 NCPA & 130\,nm rms\\
 Initial petal piston$^1$ &\,250 nm ptv\\
 Pyramid angle scatter & TBD \\
 
\hline
\end{tabular}
\footnotetext[1]{Arranged in a nuclear mode pattern.}
\end{table}

\begin{figure}[htb]
\centering
\includegraphics[width=0.65\columnwidth]{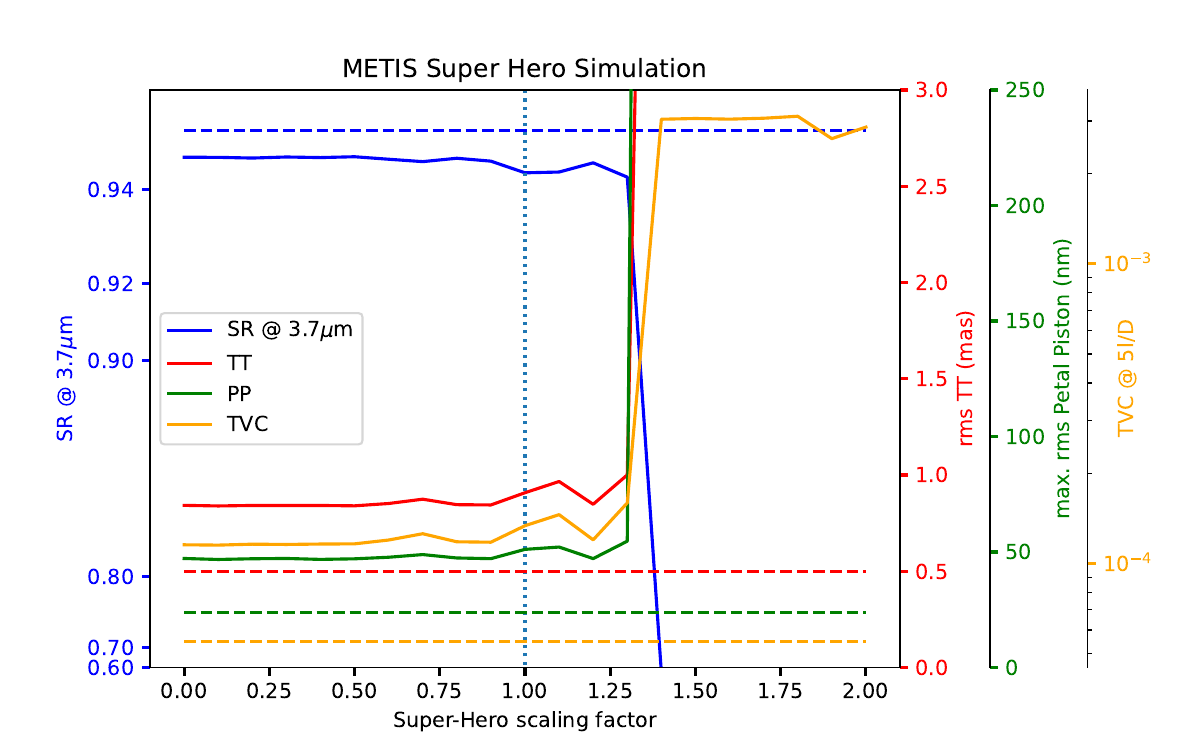}
\caption{\gls{scao} super hero performance as a function of the scaling factor affecting \gls{ncpa}s, initial petal piston, and pyramid angle deviations (c.f.\ Tab.~\ref{tab:super-hero}). The dashed lines denote the nominal performance of our baseline configuration.}
\label{fig:sim-dFull-perf-sh}
\end{figure}

In reality, an AO system will always be subject to a combination of many adverse effects, plus Murphy's law.

In order to anticipate the true performance of our system, we have combined a number of effects and introduced several "known unknown unknowns," such as modeling errors when generating the synthetic command matrices (CMs).

Tab.~\ref{tab:super-hero} summarizes the errors we have introduced into the system {\it after} calibrating for nominal conditions. That means we simulated the case where we have errors that we either do not know about during the generation of the command matrix (CM), or we do nothing about them.  We introduced two classes of errors: wind-induced tip tilt (which is always included in our simulations), a total of 21 missing M1 segments arranged in 3 flowers, and a mis-rotation and mis-alignment of M4 are always as given in Tab.~\ref{tab:super-hero}. Additionally introduced \gls{ncpa}s, a petal piston nuclear mode pattern, and an unknown angle offset to the pyramid prism faces are scaled with a common factor for analysis.  The nominal values for a scaling factor of 1 are also given in Tab.~\ref{tab:super-hero}.

Fig.~\ref{fig:sim-dFull-perf-sh} summarizes the result. All the introduced errors combined add a total of about 40\,nm rms wavefront error, making the Strehl number at 3.7\,$\mu$m drop to 94.6\%! The residual tip-tilt image motion remains submilliarcsecond, and the tip-tilt removed residual piston errors have an rms lower than 10\,nm! In summary, the system is outstandingly robust not only against individual atrocities thrown at it but also against a real-world combination of them!

\subsubsection{Force budget}
\begin{figure}[htb]
\centering
\includegraphics[width=0.65\columnwidth]{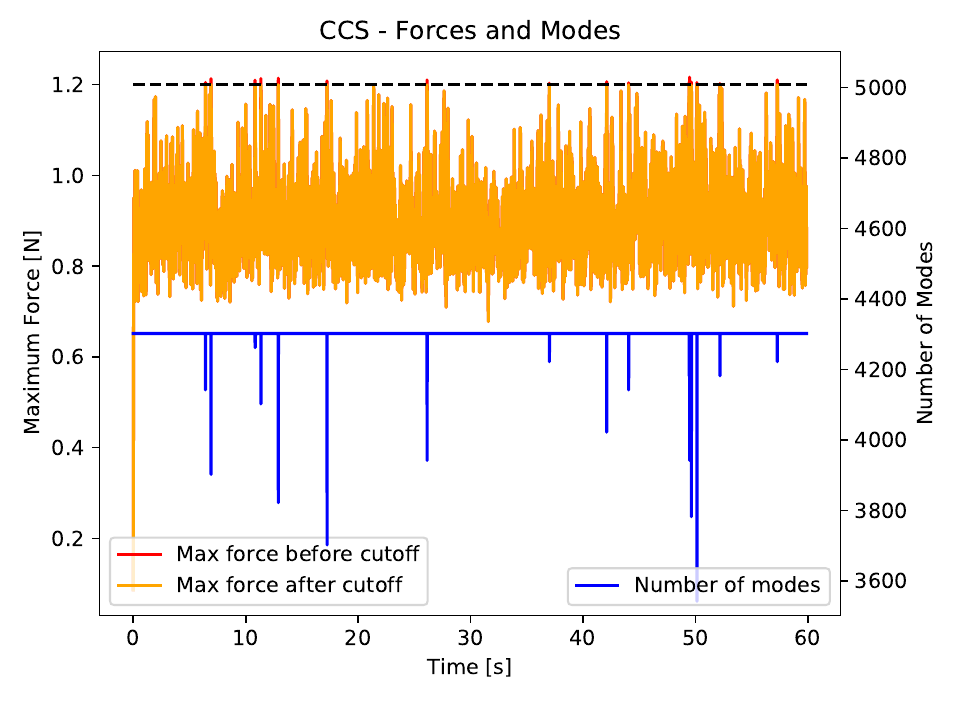}
\caption{Reduction of the number of controlled modes by the CCS whenever the calculated maximum per actuator forces exceeds the threshold of 1.2~N.}
\label{fig:sim-dFull-ccs-forces}
\end{figure}

In order to protect M4 from damage, the CCS analyses the forces that would be applied to the mirror for a given command (see Sec.~\ref{sec:baseconfig} for details). Fig.~\ref{fig:sim-dFull-ccs-forces} shows that when the calculated maximum force exceeds the threshold of 1.2~N, the CCS truncates the highest order modes until the resulting force on M4 adheres to the limit. For baseline configuration the limit was exceeded in only 0.15\% of the frames in a 60s simulation run. At no time in the simulation run the actuator gap or speed limits were broken.

\subsection{Wavefront Error Budget \label{sec:wfe-budget}}

\begin{table}[tbp]
\footnotesize
\centering

\caption[Wavfront Error Budget]{Wavefront error budget. All numbers in nm.}

\label{tab:errorbudget2}
\begin{tabular}{|l|c|} 
\hline

 \textbf{Error Sources} & Wavefront error contrib. \\ \hline 
 \textbf{AO Loop } &  \\ 
 \textbf{Error Budget}  & \\
 Fitting Error\footnotemark[1] & 107.9  \\
 Aliasing Error\footnotemark[1] & 28.8  \\
 Servo Lag Error\footnotemark[1] & 30.0 \\ 
 Photon / RON\footnotemark[1]  &  1.9  \\
 Chromatism ($z=30^\circ$)\footnotemark[1] & 8.9 \\
 Reconstruction\footnotemark[1] &  38.0 \\
 Segment Piston\footnotemark[1] &  17.0 \\ 
 regularisation and &  \\
 gain safety margin\footnotemark[2] &  25.8 \\ 
 
 \textbf{Total \gls{ao}} & \textbf{125.9} \\ \hline
 \textbf{Simulated} &  \\
 \textbf{Error Budget }  & \\
 Wind Tip-Tilt\footnotemark[3]  & 22  \\
 Wind LO Modes  & 18 \\
 Wind Load M1  & 15 \\
 M1 Static Low-Order  & 15 \\
 M1 Missing Segments  & 27 \\ \hline
 
 Residual Dispersion  & 18.4 \\
 Apodized pupil  & 45.3 \\
 Residual \gls{ncpa}\footnotemark[4] &  16.5 \\
 Pyramid prism errors &  56.6 \\ \hline
 \textbf{Total} & \textbf{153.9}\\\hline
  Total + 20\% margin & 156.9 \\ \hline
 
\hline
\end{tabular}

\footnotetext[1]{These terms have been estimated with analytic and synthetic tools.}
\footnotetext[2]{The standard configuration is not fully optimized to ensure maximum stability.}
\footnotetext[3]{This is the only contribution included in all baseline simulations. The total Wave Front Error (\gls{wfe}) up to here is 127.9\,nm rms.}
\footnotetext[4]{This includes the error contributed by the \gls{wfs} operating off-null, and the residual uncorrected \gls{ncpa}s.}
\end{table}

In addition to the two examples above, we investigated the impacts of all effects listed in Tab.~\ref{tab:errorbudget2}.  For each term listed in the lower section of Tab.~\ref{tab:errorbudget2} a dedicated simulation was set up which included the effect in question.  The \gls{wfe} determined was compared to the baseline result. and the additional error contribution was determined by quadratic subtraction of the baseline \gls{wfe} from the newly determined \gls{wfe}.  For the determination of the total \gls{wfe}, we assumed that all contributions determined in this way can be added quadratically due to an assumed independent nature of the error sources. A more elaborate approach adding several error sources and detrimental effects at once is presented in Sec.~\ref{sec:superhero}.


\section{High-Contrast Imaging \label{sec:hci}}


\subsection{HCI modes in a nutshell}

\begin{figure*}[ht]
    \centering
    \includegraphics[width=\linewidth]{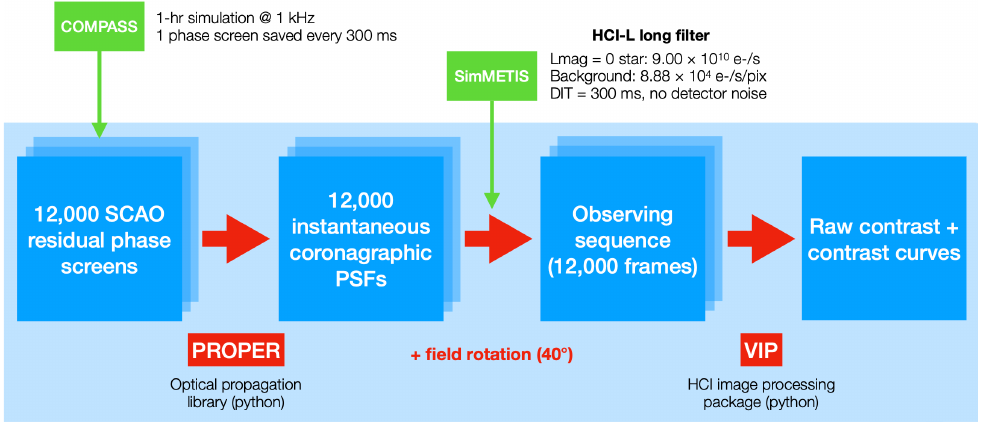}
    \caption{The High-contrast ELT End-to-end Performance Simulator (HEEPS) simulation pipeline. Credit: \cite{Delacroix+22}.}
    \label{fig:HCI_HEEPS}
\end{figure*}

METIS has two main coronagraphic modes \citep[see also e.g.][]{Brandl+22, Kenworthy+18}:
\begin{itemize}
\item Focal plane coronagraphs implemented with annular groove (vortex) phase masks followed by Lyot stops in the downstream pupil plane. They come in two flavors, the classical and the ring-apodized vortex coronagraph (CVC and RAVC resp.), and provide the smallest inner working angle of around 1 $\lambda/D$ with a 360 degrees discovery space, and good broadband characteristics ($\sim$40\% bandwith). The RAVC, implemented for $L$ and $M$-band, uses an additional ring apodizer in the upstream pupil plane to optimize the performance in the speckle-limited regime. The CVC, on the other hand, provides the highest throughput of all coronagraphic modes. CVC and RAVC are, however, very sensitive to pointing jitter or drift, and thus require active control to perform best.
\item A pupil plane coronagraph implemented with a grating vector Apodized Phase Plate (gvAPP or also simply APP) mask inserted in the pupil plane. It creates two conjugated deep dark holes providing an inner working angle of about 3 $\lambda/D$ and an outer working angle of around 20 $\lambda/D$, and provides by combining both images a close to 360 degrees discovery space. The APP provides smaller bandwidth and is less efficient in term of throughput but is also insensitive to pointing jitter and is therefore expected to be more robust against sub-optimal conditions (e.g., atmospheric or telescope vibrations). The APP will be implemented for $L$ and $M$-band, similarly to the RAVC.
\end{itemize}
The coronagraphic modes can be used to perform direct imaging in $L$, $M$ or $N$-band or combined with high-resolution spectroscopy in \textit{L} and \textit{M} bands.
The \gls{hci} requirement defined in \textit{L}-band is to reach a post-processed 5-sigma sensitivity of 3 $\times$ 10$^{-5}$ at an angular separation of 5 $\lambda/D$ (or $\sim0.1^{\prime\prime}$) and on a relatively bright star (\textit{L}$\leq$6) with a 1-hr observing sequence. This requirement was derived with the goal to enable the detection of gas giant planets and super-Earths around nearby stars.

\subsection{Simulating HCI}
To evaluate the \gls{hci} performance and verify the contrast requirement, we have developed a dedicated simulator named the High-contrast ELT End-to-end Performance Simulator (HEEPS). Because the requirement pertains to the post-processed contrast, it can only be verified by simulating a full observing sequence and by performing the appropriate data analysis. The HEEPS pipeline is illustrated in Fig. \ref{fig:HCI_HEEPS}. It includes the following steps: (i) a 1-hr SCAO simulation with COMPASS to obtain residual phase screens at a 300ms sampling, (ii) the optical propagation of the residual phase screens through a model of the METIS \gls{hci} system, taking into account any relevant environmental or instrumental effects, (iii) the simulation of a 1-hr Angular Differential Imaging (ADI) observing sequence based on the coronagraphic PSFs obtained in the previous steps, and (iv) the computation of the post-processed contrast. More details on the HEEPS pipeline and our end-to-end simulations can be found in \cite{Delacroix+22}.

\subsection{Drivers of HCI performance}

The \gls{hci} performance is influenced by a wide range of instrumental and environmental effects.  This includes effects seen by the SCAO, such as atmospheric turbulence conditions, misaligned segments in the ELT-M1, non-uniform segment reflectivity, or pupil stability, but also unseen by the SCAO such as chromatic leakage in the coronagraphic mask, pointing drifts and non-common path aberrations, and amplitude effects.
The individual and combined effects of each contributor have been the subject of detailed analyses \citep[see][]{Carlomagno+20, Delacroix+22}.
Overall, \gls{ncpa}s play a dominant role, and our ability to measure and correct them by offsetting the AO will be a key driver for the final \gls{hci} performance. In Section \ref{sec:HCI_NCPA}, we presented the expected sources of \gls{ncpa}. Here, we will discuss our mitigation strategies in detail in Section \ref{sec:HCI_NCPA_correction}. In Section \ref{sec:HCI_performance}, we summarize the expected \gls{hci} performance.


% ...




\subsubsection{Compensation of NCPA at the SCAO level}\label{sec:HCI_NCPA_correction}

As described in Secs.~\ref{sec:HCI_NCPA} and \ref{sec:DiffHOLoop}, we plan to correct \gls{ncpa}s in a dedicated auxiliary control loop.  In this section, we present the result of dedicated simulations of the impact the \gls{ncpa} correction has on the achievable contrast in METIS' \gls{hci} modes by applying offset slopes to the SCAO wavefront reconstruction.

The \gls{ncpa} phase map used in our analysis is a von Karman phase screen projected onto 100 Zernike modes, but tip and tilt, which are corrected separately by offsetting the \gls{pwfs} modulation mechanism, removed. Indeed, the focal plane wavefront sensor is expected to measure, at best, the first 100 Zernike modes.

 In line with Sec.~\ref{sec:sim-ncpa}, we consider rms levels ranging from 100 to 400 nm rms for our \gls{ncpa} phase map, representing the expected level of static \gls{ncpa} correction and water vapor seeing in $N$-band.

\begin{figure*}
    \centering
    \includegraphics[width=\linewidth]{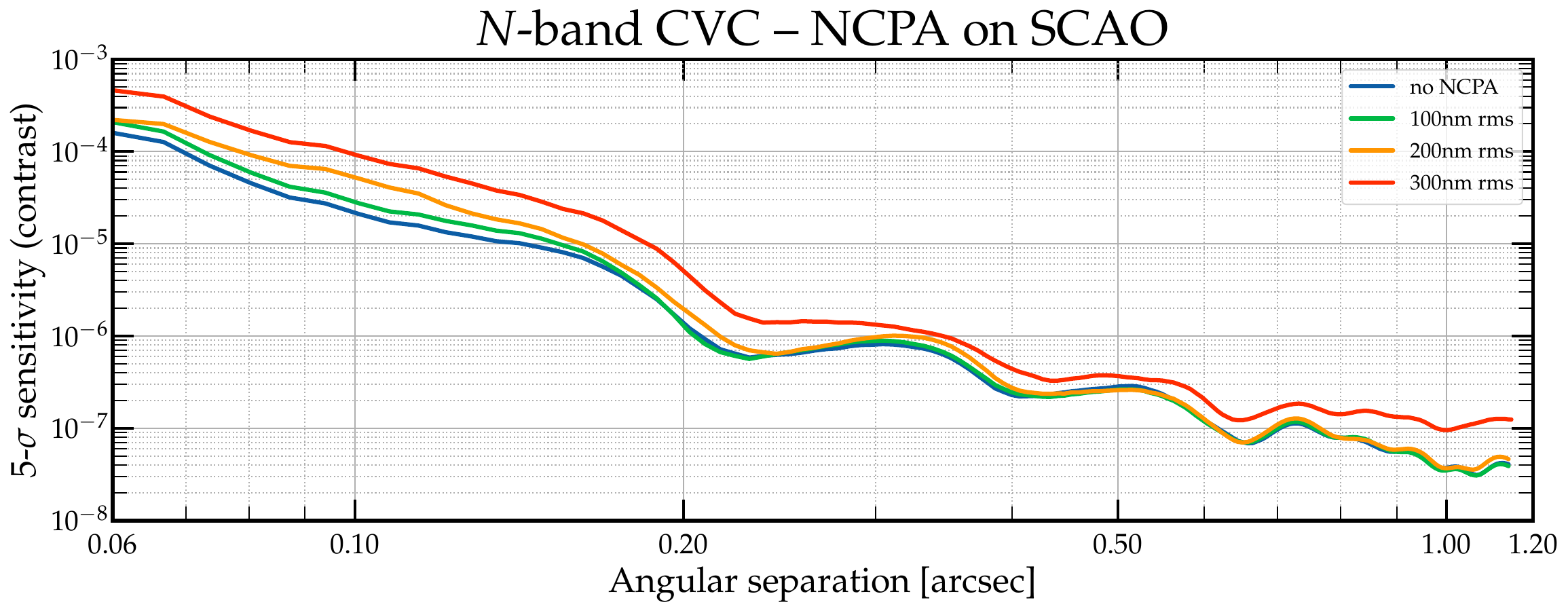}
    \caption{HCI performance in the presence of NCPA corrected by the SCAO, using a scaled von Karman phase screen. The simulations are based on 10-min SCAO sequences.}
    \label{fig:HCI_NCPA}
\end{figure*}

In Fig. \ref{fig:HCI_NCPA} we present the ADI contrast curves obtained with the $N$-band Classical Vortex Coronagraph (\gls{cvc}) mode for a magnitude $N = 0$. We use $N$-band simulations in this case because the levels of \gls{ncpa} considered here (up to 400 nm rms) will only be encountered at $N$-band in presence of water vapor seeing\footnote{The level of \gls{ncpa} correction at $L$ band will generally be of the order of 100 to 150 nm rms and driven by static and quasi-static \gls{ncpa} rather than water vapour.}. 
The 0, 100, and 200 nm rms cases show very similar performance. This result is well-aligned with the findings in Sec.~\ref{sec:sim-ncpa}. The 300 nm rms case is, however, degraded by a factor of $\sim$2.5 at 0\farcs2. 
This degradation is a direct consequence of the changes in the control parameters needed to obtain a stable SCAO system over 10-min, in particular the pyramid modulation is here 9$\lambda$/D instead of 4$\lambda$/D, and the regularisation parameter is also particularly high. However, this degradation is significantly smaller than the degradation introduced by residual water vapor seeing (about 30$\times$ for a 1Hz framerate correction, see \cite{Absil+2022}). 
For 400nm rms, we could not find a suitable set of SCAO control parameters to produce a stable 10-min simulation. It is unclear at the moment whether offloading such a high rms level will be required and it will depend on the effective amount of water vapour seeing when performing \gls{hci} observation and the fraction effectively measured by our focal plane wavefront sensor. If needed however, similarly to Pyramid wavefront sensors working at shorter wavelengths, we could track and correct optical gains to correctly apply \gls{ncpa} correction and ensure a stable control loop. 

In conclusion, trading-off SCAO performance to enable water vapor seeing correction up to at least a level of 300 nm rms should be perfectly acceptable and beneficial to the final \gls{hci} performance. 

\subsection{Summary of the expected HCI performance}\label{sec:HCI_performance}

\begin{figure*}[h!]
    \centering
    \includegraphics[width=\linewidth]{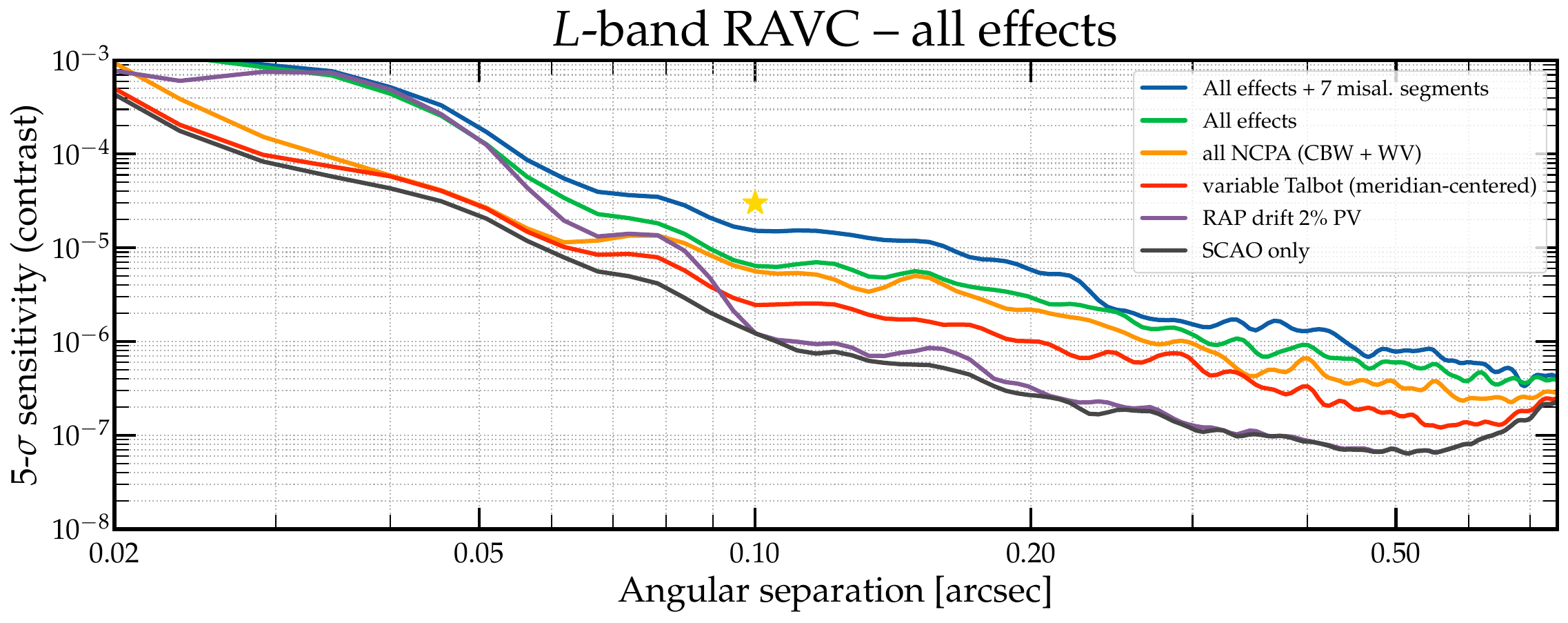}
    \caption{Post-processed contrast for the Ring-Apodizer Vortex Coronagraph in L-band. The overall \gls{hci} performance is given with and without missing segments, as well as the most prominent individual contributors. The golden star represents the \gls{hci} requirement defined in the $L$-band (3$\times 10^{-5}$ at $5 \lambda/D$). Adapted from \cite{Delacroix+22}.}
    \label{fig:HCI_performance}
\end{figure*}

\begin{figure*}
    \centering
    \includegraphics[width=\linewidth]{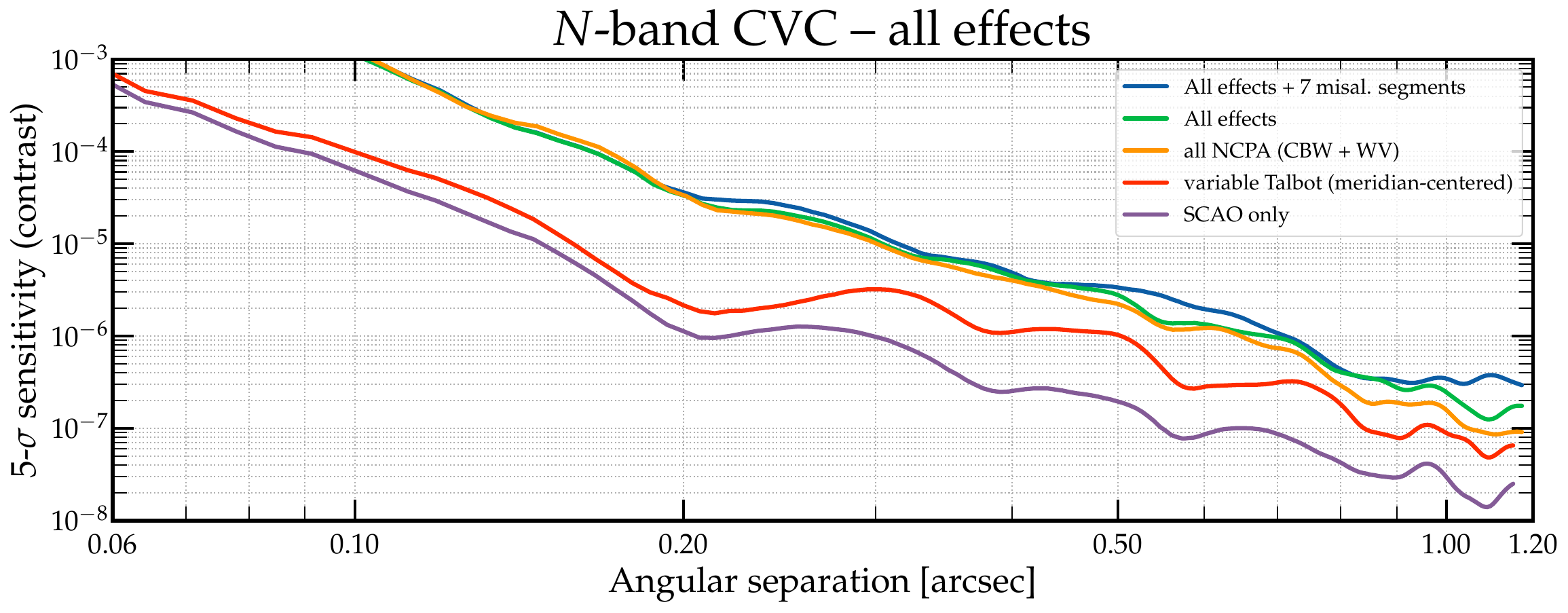}
    \caption{Post-processed contrast for the Classical Vortex Corongraph in N-band. The overall HCI performance is given with and without missing segments, as well as the most prominent individual contributors. Adapted from \cite{Delacroix+22}.}
    \label{fig:HCI_performance_N}
\end{figure*}

The expected \gls{hci} performance is obtained by assuming median seeing conditions and default SCAO parameters (e.g., wind load) and including a wide range of effects: 2\% Ring Apodizer pupil (RAP) drift, 10\% segment nonuniformity, \gls{ncpa} closed-loop residuals (considering a 1Hz repetition rate), variable Talbot effect due to chromatic beam wander, and the chromatic leakage of the vortex phase mask. More details can be found in \cite{Delacroix+22} and \cite{Carlomagno+20}. 
The final results are obtained by simulating a 1-hr ADI sequence and computing the post-processed contrast. Figures \ref{fig:HCI_performance} and \ref{fig:HCI_performance_N} show the resulting post-processed contrasts for the $L$- and $N$-bands, respectively.
Individual contrast curves are also shown for the most prominent individual contributors: 2\% pupil drift, \gls{ncpa} residuals after closed-loop, and Talbot effect.
The curves do not include however the effect of thermal background that would otherwise partly conceal the individual effects making the plots less clear. The background is, of course, appropriately weighted in when evaluating the individual contrast requirements.

The performance, see Figure \ref{fig:HCI_performance}, shows that the $L$-band requirement (\textit{i.e.} 3 $\times$ 10$^{-5}$ at an angular separation of 5 $\lambda$/D or 0.1$^{\prime\prime}$)  will be met by the RAVC in all conditions, and with a large margin in the absence of misaligned segments.
The \gls{hci} performance budget is largely dominated by \gls{ncpa} residuals, which, in $L$-band, are dominated by chromatic beam wander. The performance at $N$-band shows the severe loss due to water vapor in particular. 
We therefore expect that the on-sky performance, in particular in \textit{N}-band, will be driven by the exact level of water vapor to be corrected, our ability to measure it accurately and quickly \citep[see][]{Orban+2024}, and to offset the SCAO appropriately without decreasing the SCAO performance.

\section{Conclusions}

METIS' \gls{scao} system supports two science instruments which serve demanding science cases requiring high spectral and spatial resolution, and contrast.  Being a cryogenic instrument imposed a number of constraints that led to unconventional design decisions such as having the complete wavefront sensor module at cryogenic temperatures.  The main limitation with regard to the high contrast requirement, non-common aberrations caused by water vapor, led to the implementation of focal plane wavefront sensing applying \gls{alwfs}, where the \gls{rtc} has direct real-time access to data from the scientific focal planes. The use of a pyramid sensor at \gls{nir} wavelengths helps to overcome the petaling problem caused by the fragmented nature of the \gls{elt} pupil. 

Most components of the \gls{scao} module have already been purchased and assembled on unit level, and are currently being tested before the final integration and assembly. Comprehensive verification tests are foreseen by means of a dedicated telescope simulator.

The \gls{rtc} system also has its hardware mostly procured and assembled in the form of prototypes. The \gls{hrtc} exists as a complete prototype that includes the hard-real-time software and has already undergone performance testing regarding the timing requirements and the performance achieved in closed loop. The \gls{srtc} software is progressing well and is currently tested by means of simulations and soon also with the \gls{scao} module in combination with the telescope simulator, and potentially at the Large Binocular Telescope (\gls{lbt}).

The wavefront control concept follows an unusual approach where the reconstruction of the wavefront is logically separated from the projection onto the command mode set, and the subsequent time filtering. This approach proved exceptionally successful in avoiding the occurrence of petals, a problem abundant at the \gls{elt} where one or more of the six pupil fragments delimited by the secondary support structure spiders gets locked in a state with a constant phase offset of (an integer multiple of) one sensing wavelength. The approach also allows for an adaptation to relative offsets, rotations, or scale changes between the physical \gls{m4} of the \gls{elt} and the \gls{scao} module inside the METIS cryostat.

The high-contrast performance is also very good, reaching the required contrast of $3\times10^{-5}$ at 5\,$\lambda/D$ even when all known sources of error are applied in combination.

\backmatter





\bmhead{Acknowledgments}

METIS is built by a European consortium under the lead of the Principal Investigator institute NOVA in the Netherlands. Other consortium partners are MPIA (Germany), ATC (UK), CEA (France), ETH (Switzerland), KUL (Belgium), CENTRA (Portugal), Universities Vienna, Linz, Innsbruck (Austria), University Cologne (Germany), University of Liege (Belgium), ASIAA (Taiwan) and the University of Michigan (USA).

This project has received funding from the European Research Council (ERC) under the European Union’s Horizon 2020 research and innovation programme (grant agreement No 819155).  This research was funded in part by the NVIDIA Corporation Academic Hardware Grant Program. C.C. acknowledges support funding with DOI 2022.01293.CEECIND/CP1733/CT0012 (\url{https://doi.org/10.54499/2022.01293.CEECIND/CP1733/CT0012}) from the Portugese Fundação para a Ciência e a Tecnologia.

The authors thank Timothy Herbst for proofreading the manuscript.

\bibliography{metis-bib}

\end{document}